\documentclass[12pt]{aastex62}

\begin{document}

\parindent=1.5cm

\title{The A Supergiant Eclipsing Binary BM Cas: An Evolved, Intermediate Mass 
System}

\author{T. J. Davidge}
\affiliation{Dominion Astrophysical Observatory,
\\Herzberg Astronomy \& Astrophysics Research Center,
\\National Research Council of Canada, 5071 West Saanich Road,
\\Victoria, BC Canada V9E 2E7\\tim.davidge@nrc.ca; tdavidge1450@gmail.com}

\begin{abstract}

	The evolutionary state of the 198 day eclipsing binary BM Cas 
is examined using spectra that cover five orbital cycles. Radial 
velocities measured from SiII 6347 and SiII 6371 track the 
motion of the primary, and a mass function is found that is similar 
to that obtained by \citet{pop1977} from MgII 4481. Absorption 
from a circumsystem shell complicates efforts to measure stellar velocities 
from FeII lines. Many of the characteristics of H$\alpha$ 
emission and absorption that are associated with the shell 
vary in sync with the motion of the primary, and it is suggested that 
the shell may form from material that exits the system from L2. The 
infrared spectral-energy distribution departs from that of an A supergiant 
only at wavelengths $\geq 5\mu$m, and models are examined in which 
the secondary is obscured by an opaque envelope. 
Archived $V$ band photometry is compared with model light curves, 
and it is concluded that the A supergiant is close to filling, 
or is filling, its Roche lobe, and that the as-yet undetected 
secondary may be more massive than the primary. 
Based on the overall properties of BM Cas and its environment, 
we suggest that it is an Algol or post-Algol system, in which the A 
supergiant was originally the more massive component. If this 
is the case then the stars in BM Cas had intermediate 
initial masses. Some of the photometric characteristics 
of the primary are indicative of $\alpha$ Cyg-type variability. 

\end{abstract}

\section{INTRODUCTION}

	Many stars form with gravitationally bound companions, with 
the frequency of long lasting multiplicity increasing with mass 
\citep[e.g.][]{sanetal2012}. While the separation between 
companions can span many thousands of AU \citep[e.g.][]{sim1997}, in 
many cases stars are separated by distances that will cause them to 
interact at some point in their evolution. If such interacting 
systems eventually merge to form a common envelope (CE) then the result 
is expected to be a short period system that contains 
compact objects. Such systems have recently been detected 
in the Solar neighborhood \citep[e.g.][]{broetal2020, buretal2020}.

	While there are obvious pragmatic motivations for studying interacting 
systems with periods less than a few tens of days, longer period 
systems have the potential to sample distinct regions of parameter space, as 
they may contain components with larger masses and/or that may have experienced 
interactions at highly advanced stages of their evolution. With an orbital 
period of 198 days and an A supergiant spectral type \citep[][]{pop1977}, the 
eclipsing single line spectroscopic binary BM Cas is then a compelling target.
The light curve in the visible has a broad primary minimum that is indicative 
of components with radii that are a large fraction of the orbital separation. 
There is no obvious feature in the light curve that is associated 
with the eclipse of the secondary, indicating a substantial difference in 
the effective temperatures of the components. The spectroscopic 
signatures of the secondary might be expected to be most pronounced 
during primary minimum, when the contrast between the 
light from the components is reduced. However, neither \citet{pop1977} nor 
\citet{ferandeva1997} find direct spectroscopic signatures of this star; 
the only evidence of its presence at visible wavelengths is its gravitational 
influence on the orbit and on the shape of the primary. 

	Complex, quasi-periodic \citep[][]{ferandeva1997} photometric 
variations at the 5 -- 10\% level are also present, and these further 
complicate efforts to extract information from the light curve 
as the depth and duration of primary minimum are affected. 
Variations in the width of primary minimum are on the order of 3\%, 
while the depth can change by 0.05 mag \citep[][]{kaletal2005,kaletal2009}. 
The time of primary minimum also changes in a way that is not consistent with 
a systematic change in orbital period \citep[][]{kaletal2005,kaletal2009}, 
and this again is likely due to distortions in the light curve that are 
not related to an eclipse. The non-orbital photometric variations 
were initially thought to be due to Cepheid pulsations \citep[e.g.][]{thi1956}, 
although subsequent examination did not reveal 
Cepheid-like behaviour \citep[][]{ferandeva1997}. 

	A further complication is that radial velocity measurements are species 
dependent. \citet{pop1977} finds a $\sim 70$km/sec peak-to-peak amplitude 
difference between the velocity curves constructed from MgII 4481 and other 
lines at blue wavelengths. Moreover, he finds that, with the exception 
of MgII 4481, lines near the second quadrature are split, as might be 
expected if BM Cas were a double line spectroscopic binary, 
but this behaviour is not seen near the first quadrature. 
\citet{glaetal2008} also find species-dependent radial 
velocities, and show that the MgII 4481 line profile differs from that 
of other lines near 4470\AA. In addition, they 
measure rotational velocities, and find projected 
values of 57 km/sec from MgII 4481, and 22.7 km/sec from 
the mean of various lines. They note that, with the exception of 
MgII 4481, the lines have asymmetric profiles, and caution that the rotational 
velocities are estimates only. \citet{pop1977} considers 
the possibility that there is a circumsystem shell, but concludes that 
the spectrum is consistent with a photospheric origin. \citet{pusetal2007} 
attribute the species-dependent radial velocity 
variations to stratification and line broadening. Later in this paper 
it is demonstrated that the splitting of lines near the second quadrature 
is due to contamination from lines that form in a circumsystem shell. 

	There is no evidence for on-going mass transfer. 
\citet{ferandeva1997} do not detect the signatures of hot spots 
in UV spectra that might be expected in a 
system in which mass is being transfered. Still, 
the S/N and phase coverage of the UV spectra is not ideal. 
While the absence of strong emission lines rules out BM Cas as a system 
that is undergoing rapid mass transfer, such as in W Ser systems, 
the \citet{pusetal2007} light curve model includes a 
hot spot that has peak visibility near phases 0.78 
- 0.79. If present, such a feature could be the result of mass transfer, 
although other interpretations are possible. 

	\citet{ferandeva1997} and \citet{pusetal2007} examined 
light curve models in which the secondary is assumed to be a cool giant. 
Based largely on the width of primary minimum, these studies conclude 
that the mass ratio is $\sim 0.4 - 0.6$ (i.e. the primary is the more 
massive star) with both components filling, or close to filling, their 
Roche lobes. The orbital inclination was estimated to be 72 degrees. 

	A mass ratio less than unity and a secondary that is 
an evolved cool giant is consistent with the secondary having been initially 
the more massive star. If this is the case then the secondary shed much 
of its mass after it evolved into contact with its Roche lobe, while the 
A supergiant accreted material to become the more massive star. 
The primary has presumably evolved to the point where it will eventually 
transfer mass back to the initial donor. \citet{pusetal2007} 
suggest that BM Cas may be evolving towards a CE configuration. If this is 
correct then BM Cas is potentially an important laboratory for examining 
pre-CE evolution and the formation of compact objects in binary systems.

	While the evidence considered in previous studies does not rule out 
the secondary being a cool giant, another possibility is that it may be a star 
that is embedded in an optically thick envelope. In terms of the evolutionary 
state of the system, such a model would suggest that 
BM Cas is a post-Algol system, in which the opaque material around 
the secondary is the remnant of an accretion disk. The A supergiant primary 
was then the initially more massive star, but has evolved into a bloated 
object that is overluminous for its (now diminished) mass. IR 
emission might then be expected from the envelope that obscures the 
secondary. 

	Various properties of BM Cas are summarized 
in Table 1. BM Cas is located at a low Galactic latitude, and so it is 
subject to moderately high levels of extinction. Two E(B--V) 
estimates are provided in the table, and these indicate that the system 
is attenuated by $\sim 2 - 3$ magnitudes at visible wavelengths.

	RUWE is the GAIA renormalized unit weight error, 
which is a measure of the reliability of the astrometric quantities. 
Larger values of RUWE are indicative of less reliable astrometric 
measurements, as could result if there are discernable motions within the GAIA 
point spread function due to the relative movement of the components and/or a 
third body. \citet{penetal2022} suggest that objects with RUWE $> 1.25$ may be 
binaries in which the component motions are discerned. 
The RUWE for BM Cas in Table 1 thus suggests that the 
orbital motions of the components were not detected by GAIA, and that 
the astrometric parameters are reliable. 

\begin{deluxetable}{lll}
\tablecaption{Selected Properties of BM Cas From Past Studies}
\tablehead{Parameter & Value & Reference \\}
\startdata
$l_{II} (2000) $ & 123.296 & \\
$b_{II}$ (2000) & 1.215 & \\
 & & \\
SpT & A5Ia & \citet{pop1977} \\
 & A7Iab & \citet{bid1982} \\
 & A6Ib & \citet{ferandeva1997} \\
E(B--V) & 0.85 mag & \citet{pop1977} \\
 & 1.0 mag & \citet{ferandeva1997} \\
 & & \\
Gp & 8.485 mag & \citet{gai2023} \\
bp--rp & 1.458 mag & \citet{gai2023} \\
RUWE & 1.18 & \citet{gai2023} \\
$\pi$\tablenotemark{a} & $0.249 \pm 0.018$ mas & \citet{gai2023}  \\
$\mu_\alpha$\tablenotemark{b} & $-2.230 \pm 0.016$ mas/year & \citet{gai2023} \\
$\mu_\delta$\tablenotemark{c} & $-0.370 \pm 0.018$ mas/year & \citet{gai2023} \\
 & & \\
q & 0.4 & \citet{ferandeva1997} \\
 & $0.5 - 0.6$ & \citet{pusetal2007} \\
i & 73$^o$ & \citet{ferandeva1997} \\
 & 72$^o$ & \citet{pusetal2007} \\
f(m) & $4.05 \pm 0.10$ M$_{\odot}$ & This paper \\
\enddata
\tablenotetext{a}{Parallax}
\tablenotetext{b}{Proper motion in RA}
\tablenotetext{c}{Proper motion in Dec}
\end{deluxetable}

	Also listed in Table 1 are the mass ratio and orbital inclination 
from \citet{pusetal2007}, which is the most recent photometric study. 
However, the mass function in the table is that found in this paper. Adopting 
the mass ratio of 0.5 and the inclination of 72$^o$ from \citet{pusetal2007} 
then this mass function yields a mass of 85M$_{\odot}$ 
for the primary. In contrast, if a mass ratio of 2.0 
is assumed then the primary has a mass of 7.4M$_{\odot}$. System 
elements that are based on the examination of three light curves are presented 
in Section 6, where it is demonstrated that there are substantial uncertainties 
in the mass ratio and system inclination, both of which affect mass estimates.

	In the present study spectra of BM Cas that were recorded 
at wavelengths between 0.62 and $0.68\mu$m are combined with other 
observational information to examine the nature 
of the system. The spectra contain deep absorption 
lines of SiII and FeII that allow the radial velocity of the primary to be 
decoupled from that of a circumsystem shell, enabling a radial velocity 
curve to be obtained. The velocities are combined with 
archival $V-$band photometry to explore possible system elements. 
The characteristics of H$\alpha$, which is a prominent emission 
source in the spectrum of BM Cas and probes possible intrastellar, 
circumstellar, and circumsystem components, are also examined. The 
environment around BM Cas on the spatial scale of several 
parsecs is explored to glean additional clues into the nature of the system. 

\section{SPECTROSCOPIC OBSERVATIONS \& REDUCTIONS}

	All spectra were recorded with the McKellar spectrograph on the 
DAO 1.2 meter telescope \citep[][]{ric1968, monetal2014}. 
The spectra were acquired over 68 nights during the 2022 -- 2025 observing 
seasons, and span 6 orbital cycles. The 
spectrograph was configured with the IS32R image slicer, 
1200H grating, and 32 inch optics for all observations. The detector was the 
Site-4 CCD, which has $15\mu$m pixels in a $2048 \times 4096$ format. 
The wavelength resolution is $\sim 17000$. 

	BM Cas is a challenging target for ground-based observing 
campaigns, not only because of its long period but also because the phase 
coverage precesses by only $\sim 10\%$ between successive observing seasons.
Given these challenges, the spectra were recorded in robotic 
observing mode on nights that were scheduled with other observing 
programs that utilized the bulk of the night. The 
central wavelength of the spectrograph and the binning status of the detector 
for each night was then defined by the scientific requirements of the dominant 
program. While this results in modest run-to-run 
differences in the observational set-up, in all cases the 
central wavelength was near H$\alpha$. The wavelength coverage that is 
common to all spectra is 0.62 to 0.68$\mu$m.

	A single block of spectra were recorded on each night. 
Each block typically consisted of three 600 sec exposures 
of BM Cas that were followed immediately by an observation of a ThAr arc. 
A telluric absorption standard was also observed either immediately 
before or after BM Cas. At least one set of flat-field 
frames were recorded for each spectrograph set-up, and $10 - 20$ 
bias frames were usually recorded at the beginning and end of 
each night. 

	The exposures that were recorded on each night were combined to boost 
the S/N and facilitate the identification and suppression of cosmic rays. The 
reduction process then followed a standard sequence for single slit spectra. 
The floating and DC bias patterns were removed from the stacked frames. 
Scattered light, assessed from the region outside of the 
slit, was subtracted out. The region of the detector that contains the slit 
profile was then extracted, and the result was divided 
by the flat-field frame for that night.

	The signal in the CCD columns that fall within the 
FWHM of the slit profile of each flat-fielded 
two-dimensional spectrum were co-added to produce a single spectrum for 
that night. These were wavelength calibrated and normalized to 
a 4th order Legendre polynomial that was fit to each spectrum, with absorption 
and emission features omitted via sigma-clipping. Wavelengths near 
H$\alpha$, which is a prominent emission feature in the 
spectra, were also excluded from the fit. The final step was to apply a 
heliocentric velocity correction. 

	Two spectra that sample orbital phases near primary and secondary 
minima and that have a S/N that is more-or-less typical 
of all spectra are shown in Figure 1. Orbital phases were 
calculated using the ephemeris listed in 
Version 5.1 of the General Catalogue of Variable Stars \citep[][]{sametal2017} 
(E = 2454575.8,  P = 197.28 days), which is adopted for this paper.
There is pronounced H$\alpha$ emission while the absorption 
lines of SiII and FeII that are used for radial velocity 
measurements are also clearly visible. The insets show $15\AA$ wide 
wavelength intervals centered on the SiII and FeII lines that are 
used in Section 5 to compute radial velocities, and 
the strengths and shapes of these lines clearly differ 
between the two spectra. These differences in the line profiles are 
discussed later in the paper.

\begin{figure}
\figurenum{1}
\epsscale{1.0}
\plotone{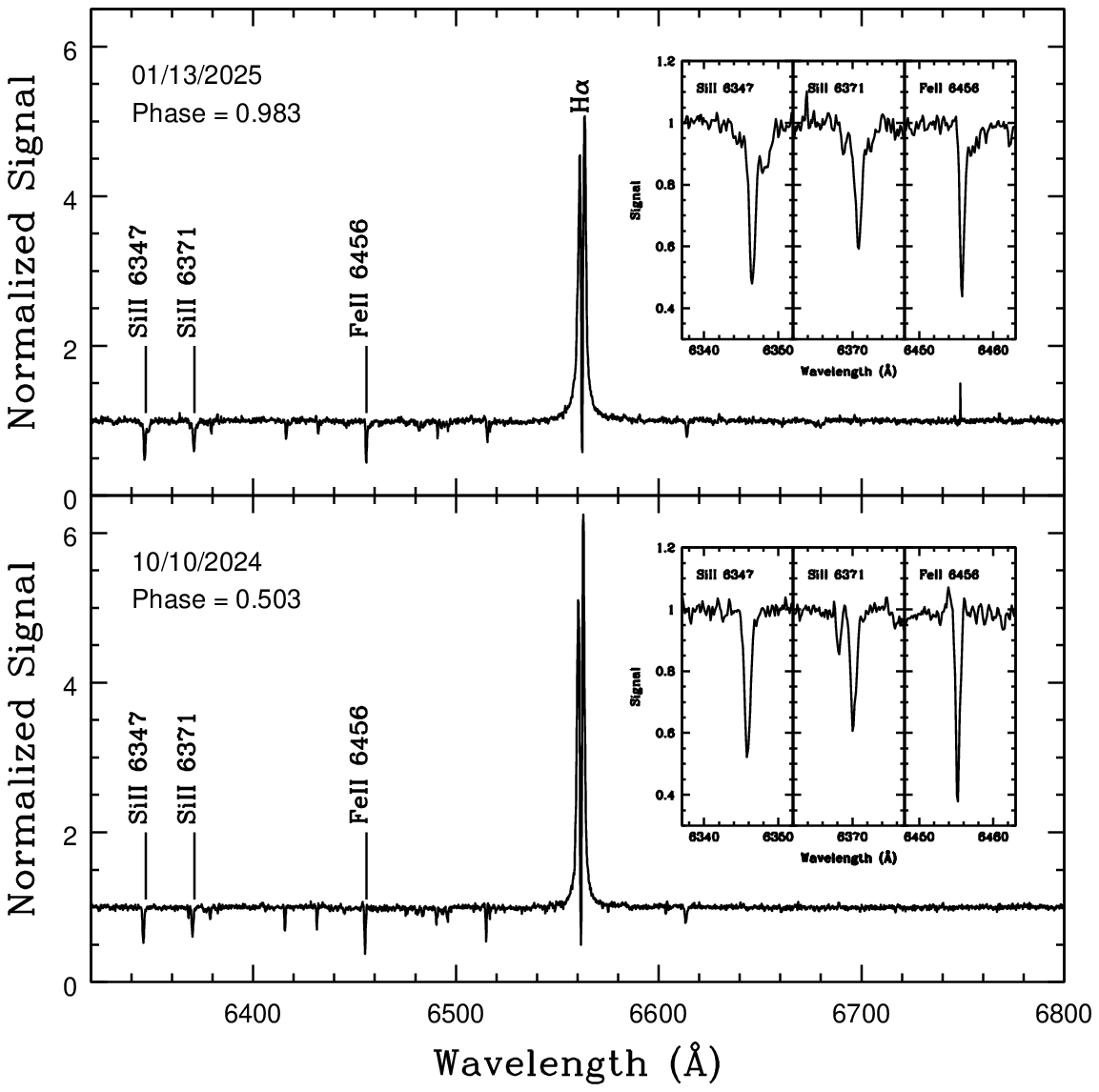}
\caption{Spectra near phases 0.0 (top panel) and 0.5 (bottom 
panel). The insets show $15\AA$ wide intervals that are centered on 
the SiII 6147, SiII 6171, and FeII 6456 lines, which are the basis for 
the velocity measurements discussed in Section 5. 
The line profiles clearly differ with orbital phase.}
\end{figure}

\section{THE INFRARED SED}

	A fundamental source of uncertainty in understanding BM Cas is the 
nature of the object that eclipses the primary. 
\cite{ferandeva1997} and \citet{pusetal2007} assume that it 
is a cool giant, and adopt T$_{eff}$ = 3500K for their 
models; stars with a substantially higher T$_{eff}$ produce a secondary minimum 
in model light curves that is not seen. If the secondary is an unobscured cool 
giant then such an object might leave a signature in the infrared (IR) 
spectral-energy distribution (SED), as it 
will contribute a progressively larger fraction of the light towards longer 
wavelengths. However, it is shown below that the $V-K$ color and IR 
SED of BM Cas are not consistent with a cool giant contributing significant 
amounts of light in the near-IR.

	The mean magnitude outside of eclipse in the published $V$ light 
curves is $V = 8.9$, while the 2MASS Point Source Catalogue 
\citep[][]{skretal2003} gives $K = 6.15$ and $J-K = 0.52$. After converting 
the 2MASS $K$ magnitude into the \citet{besandbre1988} system using 
transformation equations from \citet{car2001} then $V-K = 2.7$. 
Adopting E(B--V) = $0.92 \pm 0.07$ (the midpoint of the 
entries in Table 1), then the \citet{caretal1989} 
extinction law indicates that E(V--K) = $2.7 \pm 0.2$, and so $(V-K)_0 
= 0.0 \pm 0.1$. For comparison, the expected color for an A5 supergiant 
is $V-K = 0.5$ based on the calibration of \citet{besandbre1988}.

	Adopting effective temperatures of 8500 and 3500 K for the components, 
and assuming thay they have comparable sizes \footnote[1]{The light curve 
solutions explored in Section 7 have radii with ratios that bracket 
unity.} and radiate as black bodies then the absolute bolometric magnitudes 
of the components differ by 3.8 magnitudes. After applying bolometric 
corrections, M$_V$ differs by $\sim 6$ magnitudes. However, 
given that $V-K$ for a mid-M spectral type giant is $\sim 5 - 6$ 
\citep[][]{besandbre1988}, then the difference in M$_K$ between the primary 
and secondary is $\leq 1$ magnitude. If present, an M giant would then 
increase the intrinsic $V-K$ color of BM Cas by $\geq 0.4$ magnitudes. There 
is no evidence for such an object in the $V-K$ color estimated for BM Cas.

	The $V-K$ color estimated for BM Cas is prone to uncertainties due to 
the photometric variability of the system and the line 
of sight extinction. Still, the variations in the light curve 
introduce an uncertainty of a few tenths of a magnitude, which is not 
enough to produce a $V-K$ color that is consistent with a cool giant 
being present, while the uncertainty in the reddening account for 
$\sim \pm 0.2$ magnitude. A worse case would be if the 2MASS measurements were 
made during primary minimum, which would cause $V-K$ to be overestimated. 

	The $V-K$ color estimate is consistent with the SED at longer 
wavelengths. \citet{pop1977} compared the spectrum of BM Cas with those of 
the A supergiants HR 825 and HR 6144, and found a good match with 
the former. The infrared SEDs of BM Cas, HR 825, and HR 6144 constructed from 
entries compiled from the NASA IPAC database \footnote[2]
{https://ned.ipac.caltech.edu/} are compared in Figure 2. 
Entries at wavelengths shortward of $2.5\mu$m, 
are from the 2MASS Point Source Catalogue, 
while those at longer wavelengths are from the WISE Point Source 
Catalogue. The y axis shows instrumental magnitudes (i.e. $-2.5 
\times$ log(flux)) that have been normalized at $1.2\mu$m (i.e. in 
$J$). The SEDs have been corrected for extinction using 
the A$_V$ in the IPAC database, coupled with reddening corrections 
from \citet{yuaetal2013} for W2, and \citet{xueetal2016} for W3 and W4. 

\begin{figure}
\figurenum{2}
\epsscale{1.0}
\plotone{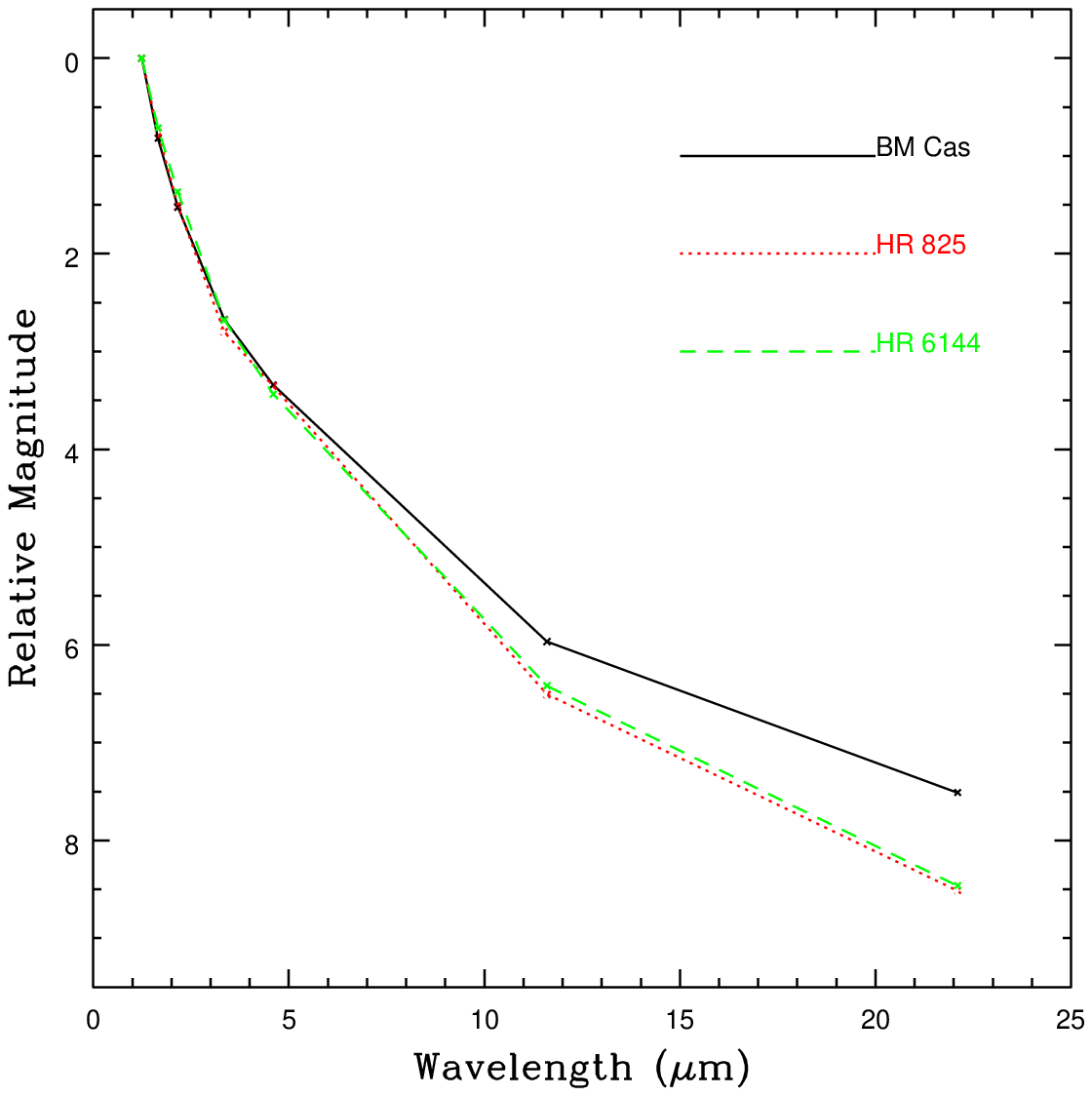}
\caption{SEDs of BM Cas (black solid line), HR 825 (red dotted line), 
and HR 6144 (green dashed line). Instrumental magnitudes 
along the y-axis have been normalized at $1.2\mu$m (i.e. in $J$). 
While the SEDs are in excellent agreement in the $1 - 2.5\mu$m interval, 
there is a $\sim 1$ magnitude excess signal in the BM Cas SED near $22\mu$m. 
As discussed in the text, this difference is larger than the expected 
uncertainties that are due to photometric variations in the light curve, the 
photometric measurements themselves, and reddening.}
\end{figure}

	The SEDs of the three stars agree in the $1 - 5\mu$m region. However, 
the SED of BM Cas shows a clear excess with respect to the other 
stars at wavelengths longward of $5\mu$m. This suggests the presence 
of a source with a T$_{eff}$ of a few hundred K, with 
no evidence for a bright cool giant. It should be emphasized that 
this excess is not a consequence of errors in the measurements. The 
measurements for BM Cas in the ALLWISE and 2MASS catalogues were 
recorded at orbital phases outside of primary minimum, thereby 
reducing uncertainties due to variations in the light curve.   
The $\sim 1$ magnitude difference between the SEDs of BM Cas and the 
a supergiants near $22\mu$m is much larger than the 
$\pm 0.1 - 0.2$ magnitude random errors expected in the WISE measurements. 
The images in the four WISE filters were also recorded at the same time, 
mitigating possible phase-related variations in the SED at these wavelengths.
Finally, uncertainties in the reddening at these wavelengths are expected to 
introduce a dispersion of only a few hundredths of a magnitude.

\section{THE ENVIRONMENT AROUND BM Cas}

\subsection{Stellar Neighbors}

	The properties of stars that are neighbors of BM Cas provide clues into 
its age and evolutionary state. Stars that are physical neighbors of BM Cas 
and that share common kinematic properties were identified using 
parallaxes and proper motions in the GAIA DR3 \citep[][]{gai2023} 
archive. The search was conducted over an angular radius of 0.085 degrees 
centered on BM Cas. This corresponds to a projected radius of $\sim 6$ 
parsecs at the distance of BM Cas, which is the 
approximate extent of the Hyades cluster. Stars with $G > 18$ were not 
considered, as the uncertainties in the proper motions climbs rapidly with 
magnitude at those brightnesses.

	Two samples of stars were extracted using 
parallax intervals that span the $1\sigma$ and $2\sigma$ uncertainties 
in the parallax of BM Cas. The $1\sigma$ sample contains stars with 
parallaxes between 0.231 and 0.267 (line of sight distances 
between 3.7 and 4.3 kpc) while the $2\sigma$ sample contains stars 
with parallaxes between 0.213 and 0.285 mas (i.e. 3.5 to 4.7 kpc). 
Objects are thus selected over a volume with a depth that 
exceeds the 6 parsec radius, which is an 
unavoidable consequence of the uncertainties in the parallax of BM Cas. 

	Additional filtering was done using proper motions. 
While radial velocities would provide additional leverage 
for identifying possible BM Cas neighbors, the velocity coverage for stars in 
the extracted region in the DR3 is sparse, and so was not considered.  
The location of stars on the proper motion plane that are in the $1\sigma$ 
and $2\sigma$ volume elements are shown in Figure 3a, where BM Cas is 
indicated with a red square. There is little difference between the 
overall distributions of the $1\sigma$ and $2\sigma$ samples in Figure 3a. 

\begin{figure}
\figurenum{3}
\epsscale{1.0}
\plotone{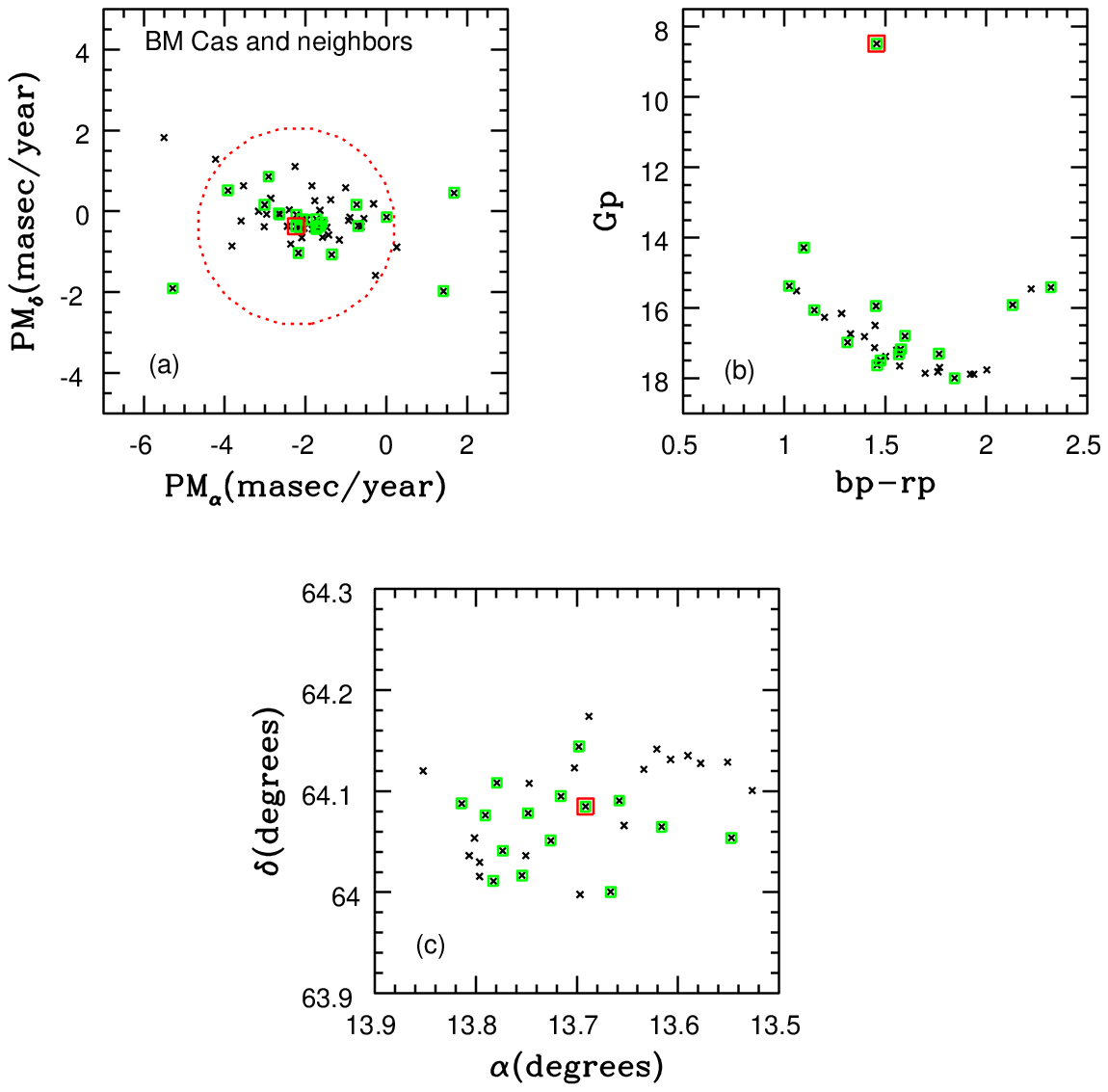}
\caption{(Panel a) Proper motions of stars with $G < 18$ that have projected 
offsets $\leq 0.085$ degrees ($\sim 6$ parsecs) from 
BM Cas. The black crosses are sources in the $2\sigma$ 
sample, while the green squares mark sources in the $1\sigma$ 
sample. The BM Cas datapoints are shown as a red square. BM 
Cas is near the center of the proper motion measurements, indicating 
that its space motions are typical for this part of the Galaxy. The circle 
marks the region of this diagram that is used to identify stars 
that were extracted to construct panels b and c. 
(Panel b) The CMD of objects that fall within the 
circle in Panel a. The photometric measurements are from 
the GAIA DR3 archive. BM Cas is much brighter than other stars in this 
sample, which is a characteristic that is common to intermediate 
mass interacting binary systems \citep[][]{dav2023}. There is an absence of 
bright blue stars that would be indicative of a population of young massive 
stars, although there are objects with $bp-rp > 2$ that are 
probable PMS stars. (Panel c) The projected on-sky distribution of objects 
with $Gp < 18$ that have distances and proper motions similar to BM Cas. 
The closest companion to BM Cas has a projected offset 
of $\sim 0.5$ parsec. The stars that are possible PMS 
objects in Panel b are in the lower right hand corner of this plot, 
and so have projected offsets of many parsecs from BM Cas.} 
\end{figure}

	The stars in Figure 3a have proper motions that span $\sim 8$ 
masec/year in RA and Dec, with BM Cas near the center of the 
data distribution. The proper motions of BM Cas are 
thus typical for its environment, and indicate that it is not a high velocity 
object. There is clustering near PM$_{\alpha} = 1.5$ masec/year and 
PM$_{\delta} = -0.5$ masec/year in both samples. 
In the absence of a physical concentration on the sky (see below) this is 
suggestive of a loose moving group.

	The main body of objects on the proper motion plane has a 
radius of $\pm 1$ mas/year centered on BM Cas and this 
area is indicated with a red dashed circle in Figure 3a. The 
color-magnitude diagram (CMD) of the sources within the red circle is shown in 
Figure 3b, where the photometric measurements are from the GAIA DR3 archive. 
BM Cas is $\sim 5$ mags brighter than the next brightest object in the 
$1\sigma$ and $2\sigma$ samples. This difference in brightness is 
a common property of intermediate mass interacting binaries 
and their surroundings \citep[e.g.][]{dav2023}. The absence of bright, 
blue stars indicates that BM Cas is not currently in an area of massive star 
formation. The four stars that have $bp-rp > 2$ have locations on the CMD that 
are consistent with them being PMS objects, although their connection 
with BM Cas is unclear given that the line of sight spans 
many hundreds of parsecs. They are also located at the edge of the 
6 parsec extraction radius.

	The projected distribution of sources on-sky 
that have parallaxes and proper motions that are similar to BM Cas is 
shown in Figure 3c. There is little or no evidence of physical clustering. 
Rather, the stellar environment around BM Cas is unremarkable, with a 
projected distance of at least $\sim 0.5$ parsec from stars with $G < 18$.
In summary, BM Cas has unremarkable kinematic 
properties, does not have bright, massive neighbors, and appears to be in 
a field environment.

\subsection{Mid-Infrared Emission}

	W1, W2 and W4 images of BM Cas, downloaded from 
the WISE \citep[][]{wrietal2010} IPAC archive 
\footnote[1]{https://irsa.ipac.caltech.edu/applications/wise/}, are shown 
in Figure 4. In Section 5 evidence is presented for a circumsystem shell 
around BM Cas. Circumsystem shells have been resolved in WISE images 
around nearby interacting binary systems \citep[][]{desetal2015, dav2022}. 
However, the FWHM of BM Cas in the WISE images indicates that if a 
luminous circumsystem shell is present then it is not resolved with the 
WISE images. Still, with the exception of the bright source in the lower 
left hand corner of the W4 image, which is in the background, there are 
no other bright MIR sources, as might be expected if BM Cas were part of a 
young star-forming region. It thus appears that BM Cas is an 
unresolved luminous, but isolated, MIR source. This is consistent with it 
being a field object.

\begin{figure}
\figurenum{4}
\epsscale{1.0}
\plotone{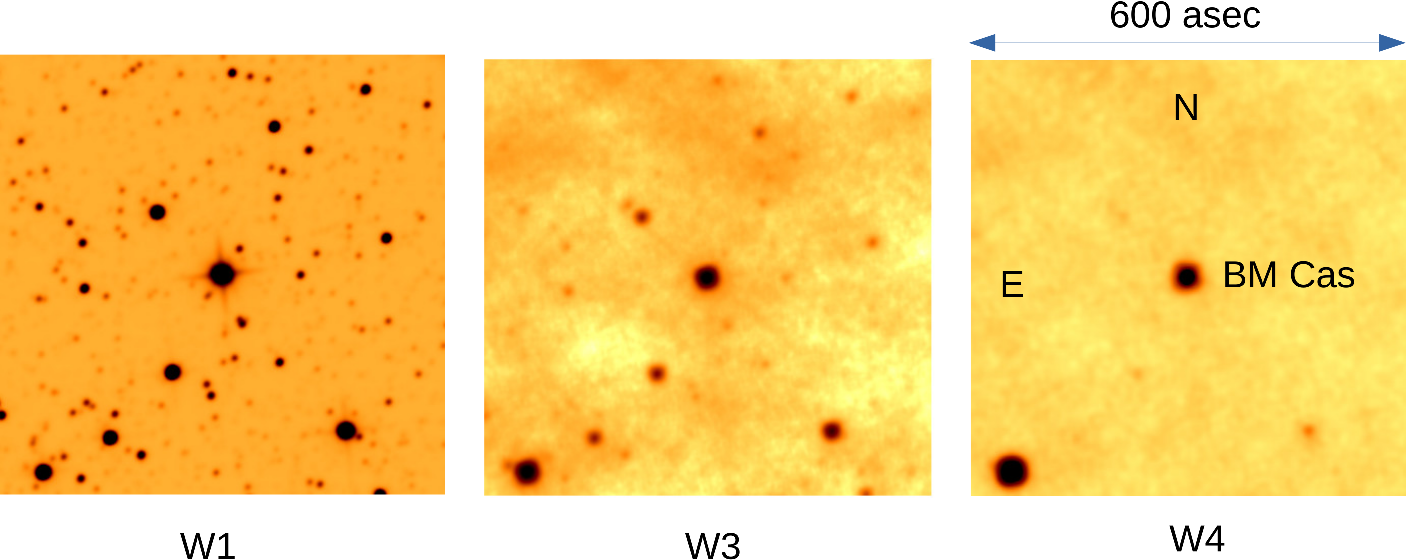}
\caption{W1, W3, and W4 images of BM Cas. Each image subtends 600 asec on a 
side, with north at the top, and east to the left. The W2 image is 
not shown as it is similar to the W1 image. BM Cas is one of the 
brightest sources in the W3 and W4 images. The bright 
source in the lower left hand corner of the W3 and W4 images is 
GAIA 524180215659014912, which has a parallax of 
0.113 asec, and so is in the background. 
The images have angular resolutions that range 
from $\sim 11$ asec FWHM in W1 to 21 asec FWHM in W4. 
BM Cas is an isolated, luminous MIR source. IR emission from 
the shell detected in the FeII lines (Section 5) is not resolved.}
\end{figure}

\subsection{The Interstellar Medium Along the Line of Sight}

	The Canadian Galactic Plane Survey \citep[CGPS;][]{tayetal2003} 
mapped interstellar emission at 408 and 1420 MHz over a large swath of 
the Northern Milky Way. IRAS \citep[][]{neuetal1984} images 
are included with the radio maps to allow the 
distribution of warm dust also to be examined. 
The projected on-sky distribution of emission near BM Cas from 
CGPS maps is examined in Figure 5, where $0.6 \times 0.6$ degree 408 MHz, 
CO, and HI CGPS maps are shown, along with IRAS 60 and 
$100\mu$m images. BM Cas is a distinct source in the CGPS 
$10\mu$m and $25\mu$m IRAS images, although these are not shown given 
the superior angular resolution of the WISE images in Figure 4.

\begin{figure}
\figurenum{5}
\epsscale{1.0}
\plotone{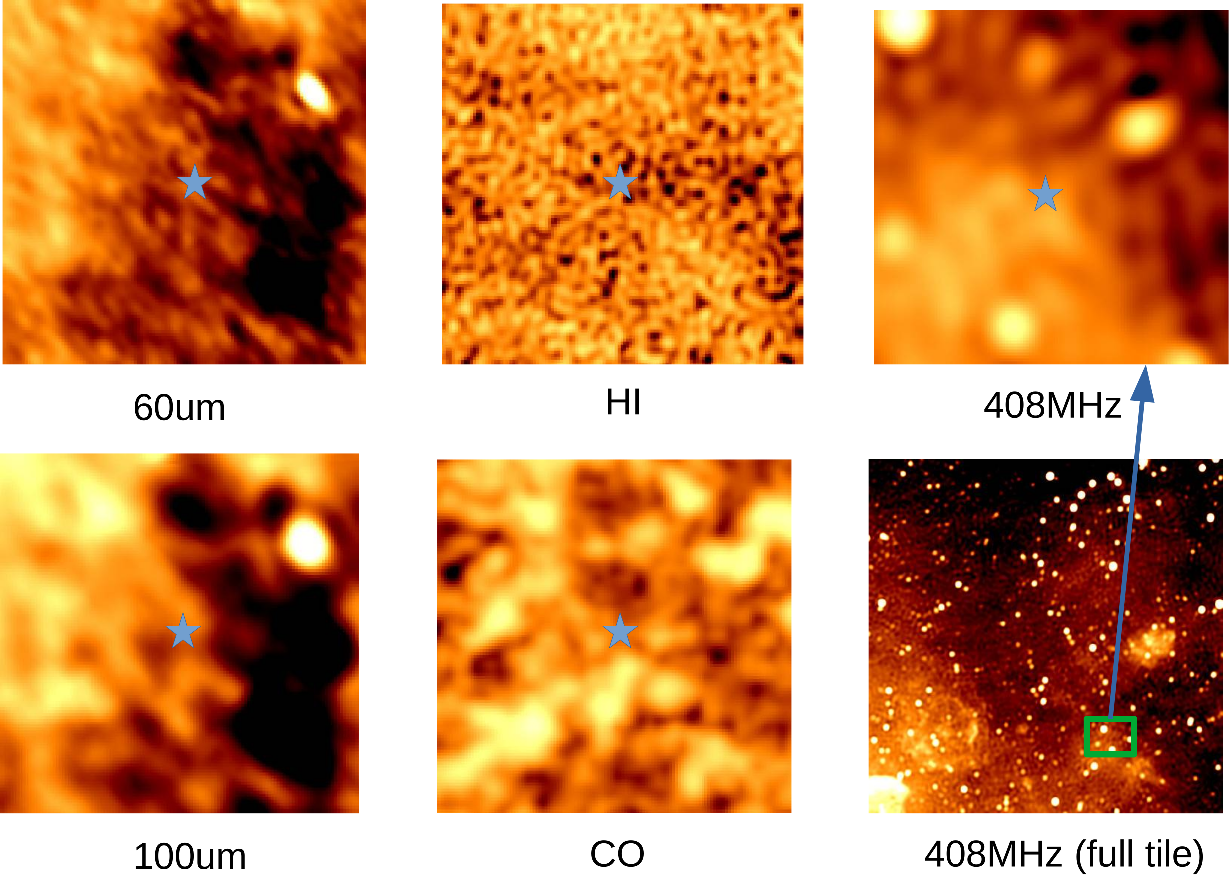}
\caption{Interstellar emission along the BM Cas line of sight. With the 
exception of the 408 Mhz image in the lower right 
hand corner, the images cover $0.6 \times 0.6$ degrees. 
North is at the top, and East is to the left. The location 
of BM Cas is indicated with a star. The HI and CO data in the CGPS cover a 
range of wavelengths, and the data cubes have been collapsed to produce an 
integrated signal along the line of sight. A 408 MHz image that covers a larger 
area is also shown to enable a search for large scale structures in the ISM 
near BM Cas. While BM Cas is located in an area of moderate emission from 
gas and hot dust, the line of sight passes through the Galactic disk, 
and so depth effects are significant.}
\end{figure}

	The CO and HI channel maps in Figure 5 have been collapsed along the 
velocity axes to show an integrated picture of all emission sources 
for consistency with the $60\mu$m, $100\mu$m, and 408 MHz images. A wide 
field 408MHz image is also included in Figure 5 to search for large-scale 
structures such as SNRs and bubbles. A bubble is present 
over a degree to the east of BM Cas; otherwise, there is an absence of 
remarkable features in the ISM. In summary, while there is emission 
from atomic, molecular, and dust sources along the BM Cas sight line, 
this emission is not remarkable when compared with that in the 
surrounding areas.

\section{RADIAL VELOCITIES}

	Radial velocities were measured from the SiII 6347, SiII 6371 and 
FeII 6456 lines, which are the deepest metal lines in the DAO spectra. 
Measurements were made in the line cores, and so gauge the velocity 
of the deeper component of the SiII doublets. A cross-correlation technique 
spanning a range of wavelengths was not applied given the species-specific 
nature of the velocities, as found by \citet{pop1977}. 

	Phased velocity curves are shown in Figure 6. These 
are consistent with a circular orbit, in agreement with \citet{pop1977}. 
A circular orbit is noteworthy given that the frequency 
of eccentric orbits among binary systems increases with orbital 
period \citep[e.g.][]{hei1969}. Interactions between components and with 
the surrounding medium could circularize an initially eccentric orbit. 

\begin{figure}
\figurenum{6}
\epsscale{1.0}
\plotone{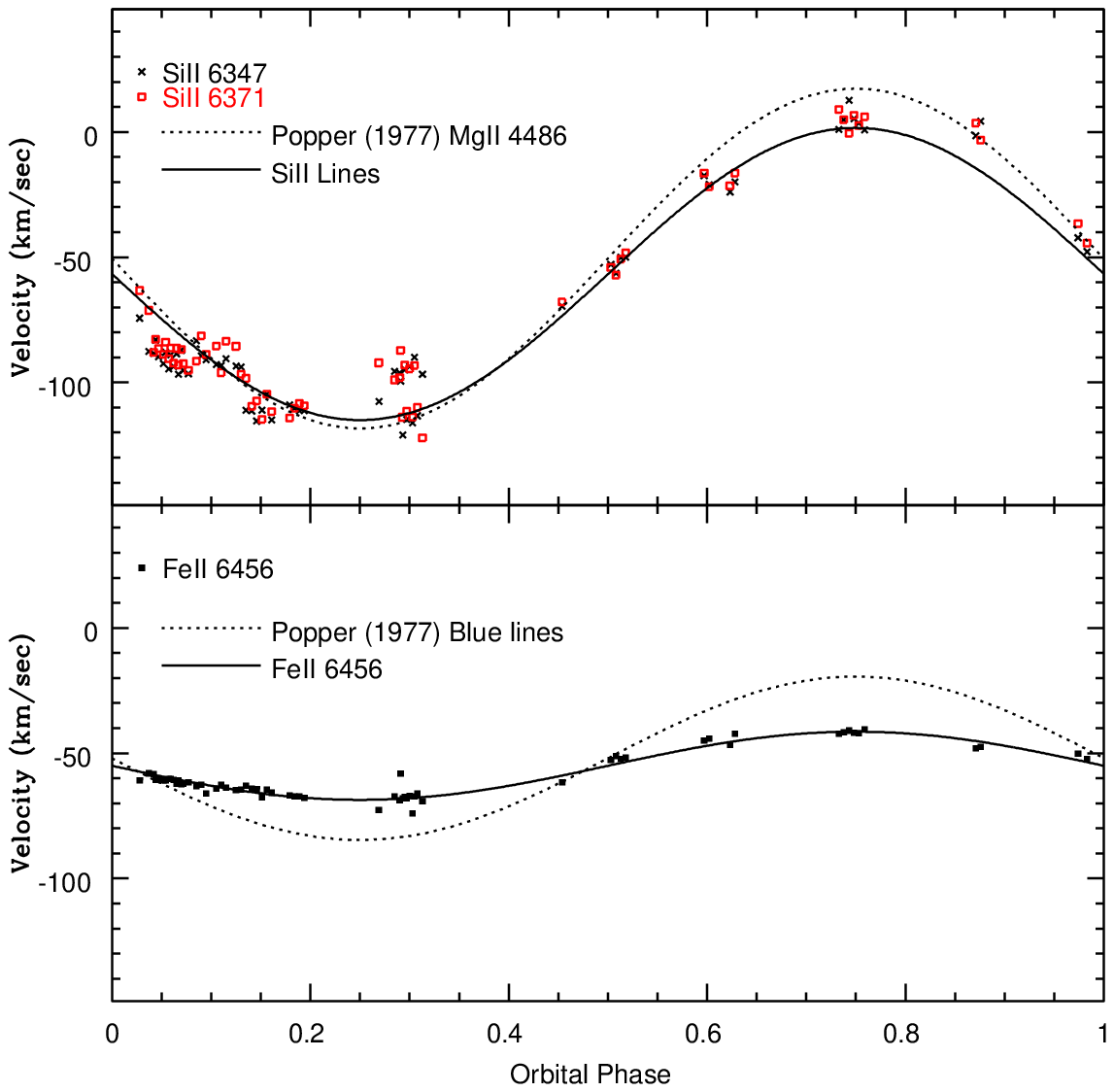}
\caption{Phased velocity curves obtained from the cores of SiII 6347, SiII 
6371, and FeII 6456. The scatter in the FeII measurements at a given phase 
suggests a measurement uncertainty of $\pm 1 - 2$ km/sec, while the scatter 
among the SiII measurements suggests a $\pm 5$ km/sec uncertainty. 
A cycle-to-cycle dispersion in the SiII and FeII 
measurements is seen near phase 0.3. The solid lines are 
fits to the measurements that assume a circular orbit. 
Velocities from the SiII lines were averaged together to generate 
the SiII curve. Velocity curves based on measurements made 
by \citet{pop1977} at shorter wavelengths are shown as dotted lines for 
MgII 4481 (top panel) and for other lines at blue wavelengths (bottom 
panel). A table that contains the SiII and FeII radial velocities 
plotted in this figure is available as 'data behind the figure'.}
\end{figure}

	There is generally good agreement between the 
velocities obtained from SiII 6347 and 6371, and the dispersion 
in the curve suggests measurements errors of a few km/sec. Cycle-to-cycle 
differences at the 10 - 20 km/sec level are seen 
near phase 0.3, and \citet{pop1977} also reported peculiarities in 
velocity measurements near this phase. The FeII 6456 velocities 
also define a tight curve, with increased scatter in the 
FeII velocities near phase 0.3. Still, cycle-to-cycle 
variations in the FeII 6456 measurements are smaller than those among 
the SiII lines. The overall dispersion in the FeII velocities 
suggests measurement uncertainties of $\pm 1 - 2$ km/sec. 

	Curves that assume a circular orbit were fit to the measurements, 
and these are compared with the observations in Figure 6.
The velocity curve amplitudes and system velocities ('$\gamma$') are listed 
in Table 2. The SiII and FeII lines yield system 
velocities that agree with those found by \citet{pop1977}. 
The radial velocity curve half amplitudes are $58.3 \pm 1.4$ km/sec 
for the SiII lines, and $13.6 \pm 0.4$ km/sec for FeII 6456. 
The mass functions are then $4.05 \pm 0.10$ (SiII) and $0.051 \pm 0.002$ 
(FeII) M$_{\odot}$. The differences between these mass functions have enormous 
implications for determining the component masses.

\begin{deluxetable}{llc}
\tablecaption{Velocity Curve Measurements}
\tablehead{Species & Half Amplitude\tablenotemark{a} & $\gamma$\tablenotemark{a} \\
 & (km/sec) & (km/sec) \\}
\startdata
SiII 6342, 6371 & $58.3 \pm 1.4$ & $-56.7 \pm 0.9$ \\
FeII 6456 & $13.6 \pm 0.4$ & $-55.0 \pm 0.3$ \\
H$\alpha$ absorption & $1.5 \pm 0.1$ & $-52.4 \pm 0.1$ \\
H$\alpha$ emission (narrow) & $5.5 \pm 0.9$ & $-51.7 \pm 0.6$ \\
\enddata
\tablenotetext{a}{Assumes a circular orbit, with motions that are in phase 
with the primary. $1\sigma$ uncertainties are shown.}
\end{deluxetable}

	\citet{pop1977} and \citet{glaetal2008} found that 
velocities measured from MgII 4481 differ from those obtained from 
other lines at blue wavelengths, with the amplitude of the 
MgII 4481 velocity variations being greater than that defined by combining 
results from other lines. The FeII and SiII 
velocity curves obtained from the DAO spectra reveal a similar situation. The 
SiII velocities are broadly consistent with those obtained by \citet{pop1977} 
for MgII 4481, although there is a large discrepancy near 
phase 0.3, where some of the SiII velocities depart from the MgII 
curve by $\sim 20$ km/sec. In contrast, the amplitude 
of the FeII velocity variations is smaller than the curve defined by 
the majority of blue lines in the \citet{pop1977} spectra.

	There is clear sub-structuring in the FeII 6456 line, 
and the characteristics of the sub-components provide clues into the 
differences between the SiII and FeII velocities. 
The structure of the FeII line profile is examined in 
Figure 7, where the means of spectra centered 
on the SiII and FeII lines between phases 0.15 and 
0.25 are compared with the means of spectra between phases 0.65 and 0.75.
The FeII lines in the DAO spectra have two distinct components: 
(1) a narrow component that is deeper than the SiII lines, which 
has a wavelength motion that is much smaller than that of the SiII lines, 
and (2) a shallower, possibly broader, component with a depth that is 
comparable to that of SiII 6371 and that moves in approximate agreement with 
the SiII lines. \citet{glaetal2008} found asymmetric 
line profiles among FeII and TiII lines near 4470\AA\ that are similar 
to those in the DAO spectra.

\begin{figure}
\figurenum{7}
\epsscale{1.0}
\plotone{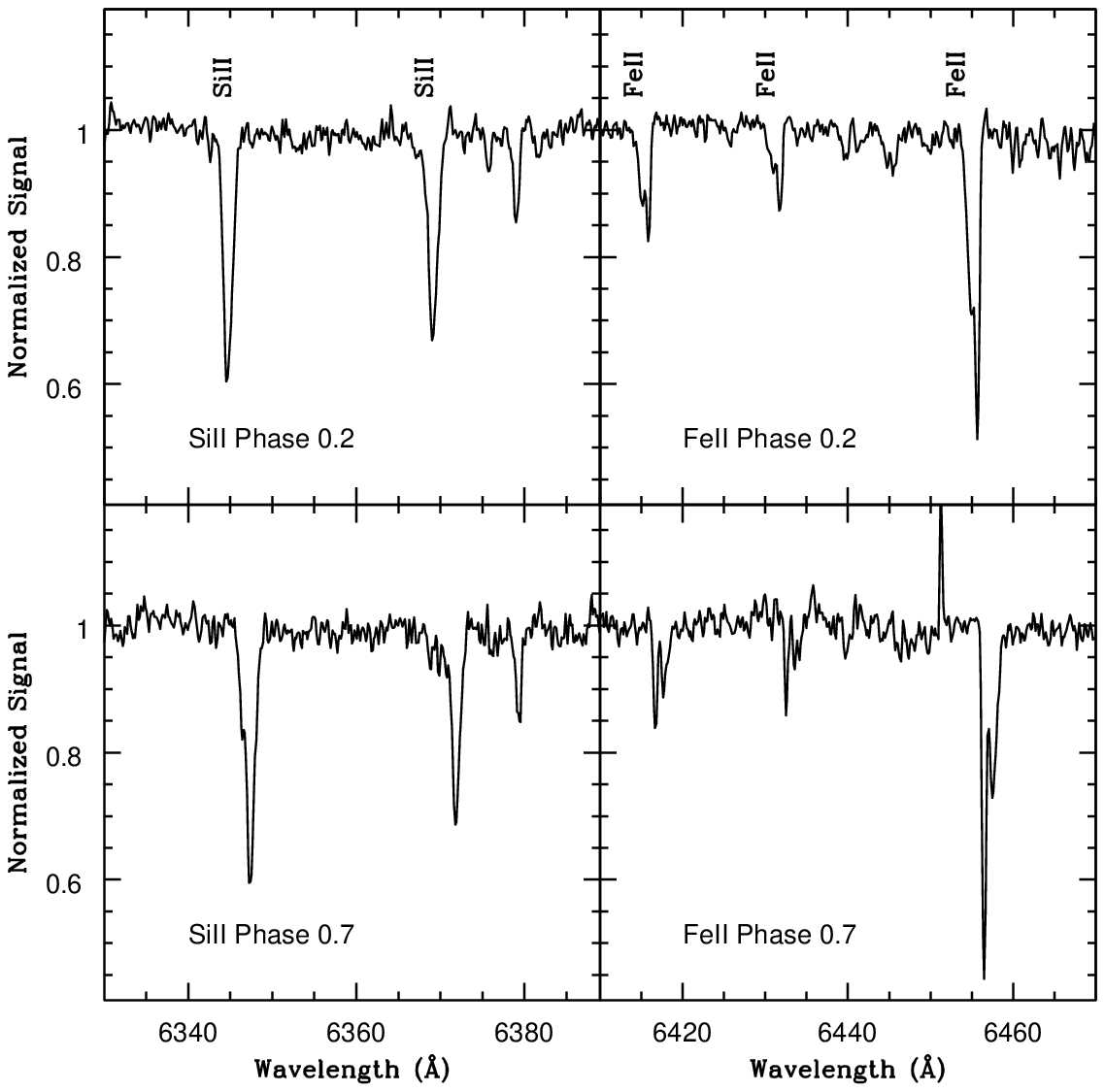}
\caption{SiII and FeII lines in mean spectra at phases 
0.15 - 0.25 (top row) and 0.65 - 0.75 (bottom row). The profiles of the 
SiII and FeII lines clearly differ. Whereas there is only modest sub-structure 
in the SiII lines, there are two distinct components in the FeII lines. 
The weaker component is likely stellar in origin while the 
deeper component has characteristics, such as a narrow width, 
that are suggestive of a circumsystem shell. 
The FeII components are cleanly split near phase 0.7, 
and similar behaviour was noted among many of the lines at blue wavelengths 
examined by \citet{pop1977}; the exception was MgII 4481.}
\end{figure}

	Given its width and modest velocity variation, it is 
likely that the narrow FeII component originates in a circumsystem 
shell, and thus does not track the motion of either star. While there is 
evidence for sub-structuring in the SiII profiles, the contribution from 
the shell component is modest when compared with the main body of 
the line, with a corresponding reduced effect on the velocity measurements.
We suspect that shell features may be present in many of the lines examined by 
\citet{pop1977}, who found that, with the exception of MgII 4481, lines 
in his spectra showed splitting near phase 0.75, but not near phase 
0.25. Splitting of the FeII profile near phase 0.75 is seen in Figure 7.

	Given the presence of a likely shell contribution in the 
FeII line profiles, we argue that the SiII lines and MgII 
4481 are likely a more reliable tracer of the primary star's motion than 
the deep component of the FeII lines. We do not consider the shallow FeII 
component for velocity analysis because it is heavily blended with the shell 
component for much of the orbital cycle. Still, we estimate 
that the half amplitude of the shallow FeII component is $\sim 60$ km/sec, 
which is in rough agreement with that obtained from SiII 6347 and SiII 6371.

	The SiII 6347, 6371 and MgII 4481 lines share 
an excitation pedigree that is based on UV resonance transitions 
\citep[e.g. discussion in][]{armandsho2022}, and so similarity in their 
kinematic properties may not be surprising. These lines also may 
not be photospheric in origin, but might originate in a disk around the 
primary. If this is the case then they should still serve as a proxy for 
tracking the motion of the primary, although stratification might produce
line to line velocity differences, as suggested by \citet{pusetal2007}.

	How do the velocities measured from SiII 6347 and 6371 compare 
with those obtained from other SiII lines? \citet{pop1977} concludes that SiII 
4128 and 4130 have depths that are similar to those of classification 
standards, and then argues that they likely do not have a shell 
origin. While the extent to which shell spectra may deviate 
from those of classification standards is not always obvious, the blue 
SiII lines were almost certainly included in the 17 line measurements that 
\citet{pop1977} combined to obtain velocities. Although the resulting velocity 
curve has a smaller amplitude than that produced from SiII 6347 and 6371, 
this does not necessarily mean that there is an inconsistency in the kinematic 
behaviour of the blue and red SiII lines, given that the \citet{pop1977} 
velocity curve is the result of combining velocities from a number of 
species. In fact, \citet{pop1977} noted that there is large scatter among the 
17 velocities that were combined. However, with the exception
of MgII 4481, he did not discuss systematic differences 
between lines. In summary, it is not clear to what 
extent the velocities obtained from SiII 4128 and 4130 differ 
from those obtained from SiII 6347 and 6371.

\section{H$\alpha$}

	H$\alpha$ is the dominant feature in the 1.2 meter spectra, 
and its characteristics provide clues into the circumstellar and 
circumsystem environment of the system. The morphology of H$\alpha$ in 
the BM Cas spectrum is examined in Figure 8, where mean profiles 
at phases 0.1 and 0.5 are compared. The overall strength of 
emission at these phases differs, as do the relative strengths of the 
peudo-peaks on either side of the absorption feature. 
Substructure is present in the emission profile, and the morphology changes 
with phase. 

\begin{figure}
\figurenum{8}
\epsscale{1.0}
\plotone{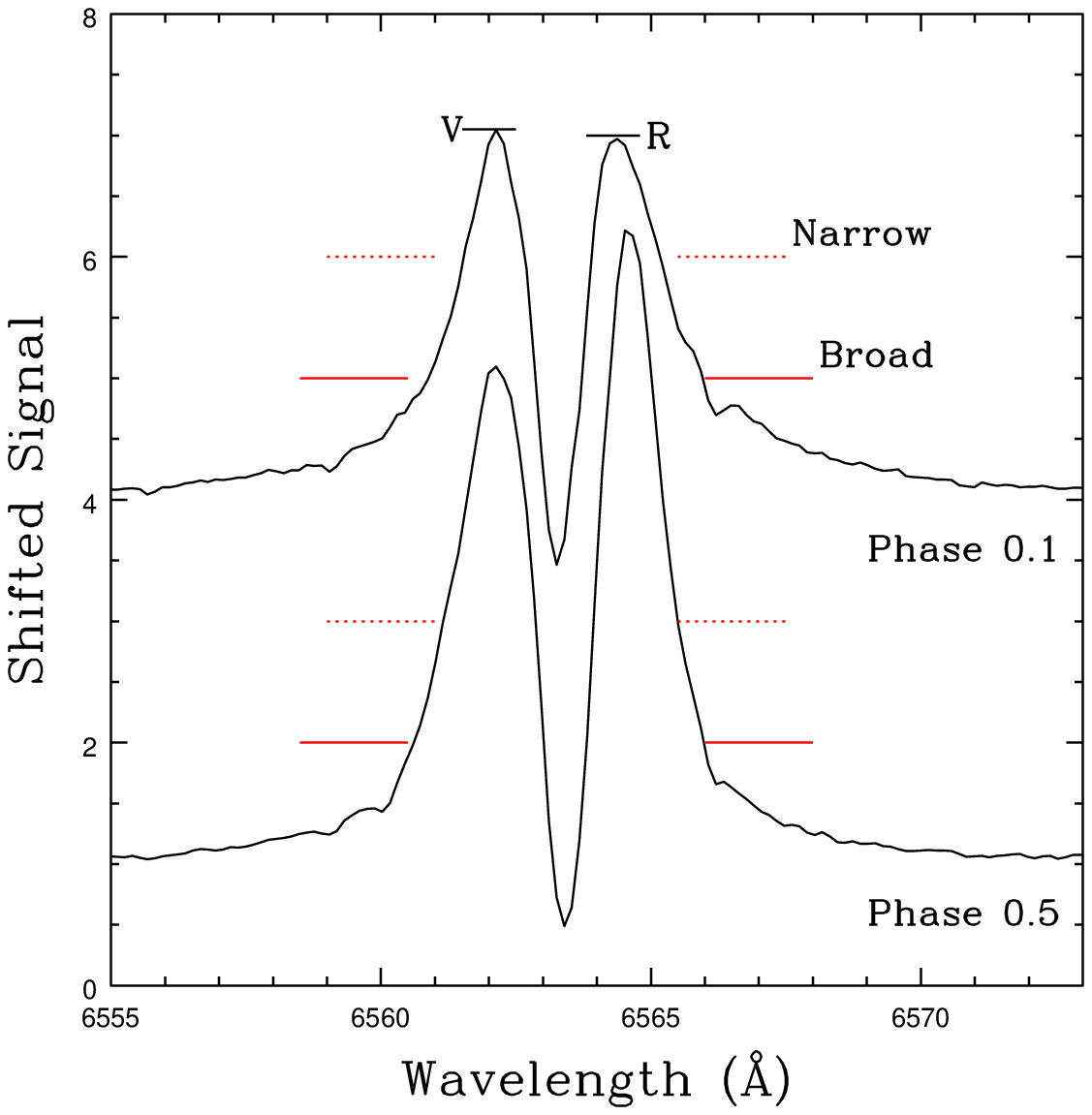}
\caption{H$\alpha$ profiles extracted from the mean of spectra in the phase 
intervals 0.05 -- 0.15 (top spectrum) and 0.45 -- 
0.55 (bottom spectrum). The locations of the V and R measurements 
are indicated for the phase 0.1 profile. The points in the line profile 
where the velocity measurements were made are marked with the solid 
(broad component) and dashed (narrow component) red lines. 
The height and shape of H$\alpha$ both change with phase. 
The red edge of the phase 0.1 emission profile bulges 
out near its base, indicating that there is an emission source that is 
receding from the system. This feature is less pronounced in the phase 0.5 
profile.}
\end{figure}

	The equivalent width and the central 
velocity of the absorption component, as well as the 
velocity of the emission component and the relative heights of the 
pseudo-peaks that bracket the absorption, were measured. The 
orbital phase-dependence of these measurements is examined in Figure 9. 
The point with the largest equivalent width in the top panel is from 
the September 4, 2023 spectrum. While this spectrum has a relatively low S/N, 
the other H$\alpha$-related measurements made on that night do not stand out.

	The V and R measurements peak between 
phases 0.45 and 0.65. This suggests that the emitting 
region and/or the source of exciting radiation are/is not 
uniformly distributed. V/R also changes with phase, indicating that the shape 
of H$\alpha$ and the height of the two pseudo-emission peaks, which 
in the absence of prominent transient substructures are shaped by the 
relative central wavelengths of the absorption and emission components, also 
change with phase. 

\begin{figure}
\figurenum{9}
\epsscale{1.0}
\plotone{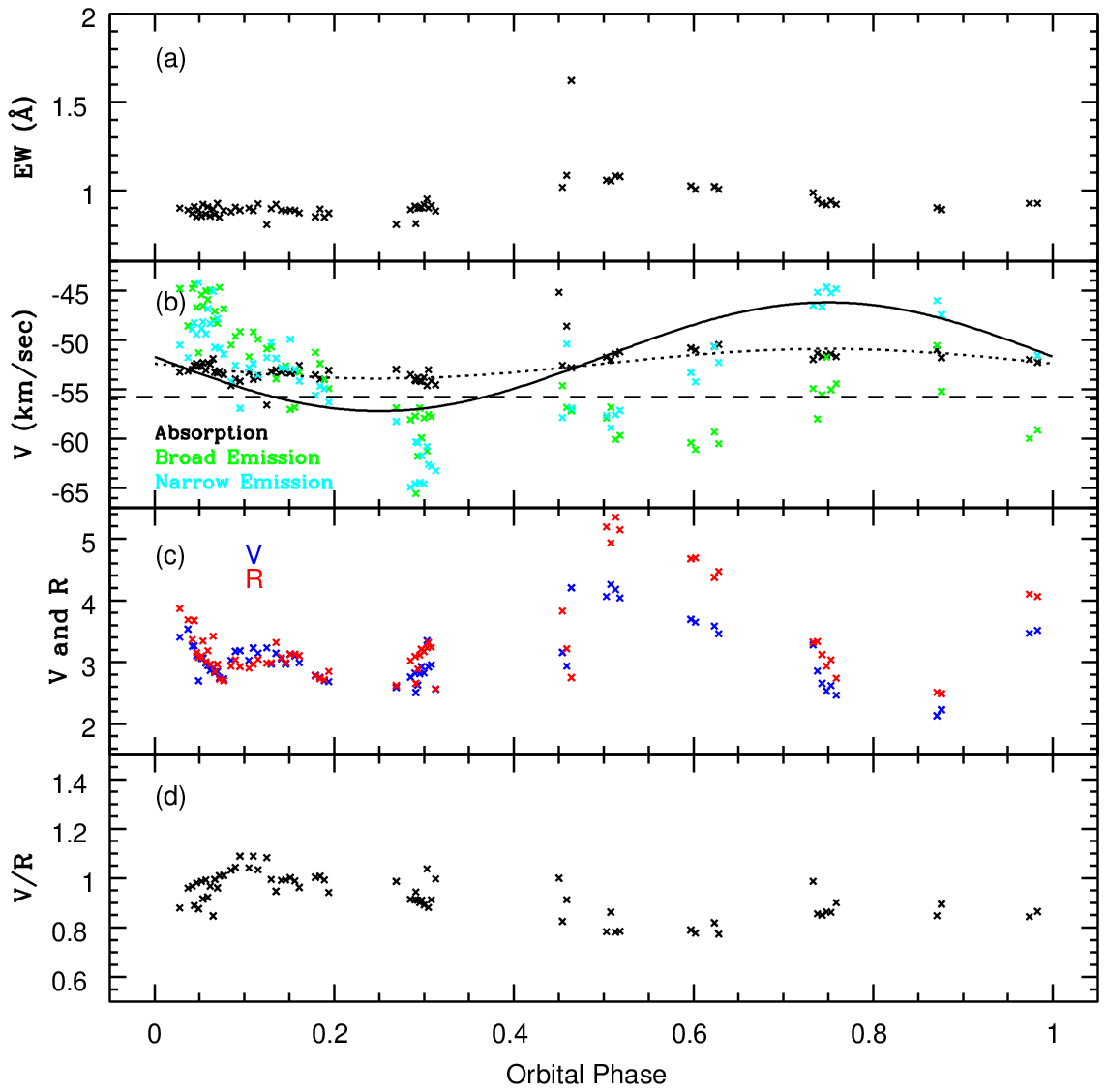}
\caption{Phase related characteristics of H$\alpha$ absorption and emission. 
(Panel a:) The equivalent width of the absorption component. 
(Panel b:) Heliocentric velocities of the absorption and emission 
components. The dashed line is the mean of the $\gamma$ 
velocities measured from SiII and FeII in Table 2, while the dotted line is 
a fit to the H$\alpha$ absorption velocities assuming a circular orbit 
that is in phase with that of the primary. The velocity of 
H$\alpha$ absorption varies with orbital phase in approximate sync 
with the motion of the primary. The solid line is a fit to the 
narrow emission velocities assuming motion that is synced 
with the primary. There is substantial scatter about the fit, 
with a possible phase lag with respect to the motion of the primary. 
The motion of the broad emission component is clearly out 
of sync with that of the primary and this is likely 
due to a kinematically distinct H$\alpha$ component that appears at some 
phases (see text). (Panel c:) The heights of the pseudo-peaks 
that bracket the absorption line, which vary 
with orbital phase, and are strongest between phases 0.45 and 0.65. (Panel d:) 
V/R, which also varies with orbital phase. The variation in V/R is likely due 
in part to the relative motions of the absorption and emission components. 
An ASCII table that contains the radial velocities and V and R measurements 
plotted in this figure is available as 'data behind the figure'.}
\end{figure}

	The velocity and equivalent width of H$\alpha$ absorption 
were found by fitting a Gaussian to the absorption profile. 
Both quantities vary throughout the orbital cycle, with the equivalent 
width reaching a maximum between phases 0.45 and 0.65. Such a variation 
in the equivalent width with phase might be expected, as the equivalent width 
should be greatest near secondary minimum when the primary 
is not obscured by the secondary. The velocity of H$\alpha$ 
absorption varies in phase with the motion of the primary, further 
suggesting a causal connection between H$\alpha$ absorption and 
that star. $\gamma$ and the half amplitude of the absorption velocity curve, 
where the former assumes that the absorption velocities 
vary in phase with the primary, are listed in Table 2. The amplitude 
of the H$\alpha$ absorption velocity curve is a few km/sec. 
$\gamma$ measured from H$\alpha$ absorption is significantly larger than the 
$\gamma$ measurements made from the SiII and FeII lines.

	The midpoint of the H$\alpha$ emission profile has been 
used in other studies to trace the motions of Be stars in 
binary systems \citep[e.g.][]{ruzetal2009}. \citet{wanetal2023}
demonstrate that well-defined radial velocity curves can 
be extracted when the H$\alpha$ profile does not show complicated variations. 
The central wavelength of the BM Cas emission profile, and hence its 
radial velocity, was estimated by taking the mean of wavelengths in the 
shoulders of the profile at normalized intensities that are $2\times$ and 
$3\times$ higher than the continuum. The measurements at the lower normalized 
intensity probe the mean velocity near the base of 
H$\alpha$ (the 'broad' component), while the measurements at the higher 
intensity monitor the velocity closer to the peak of the emission (the 'narrow' 
component). The resulting velocity curves are shown in Figure 9.

	The broad and narrow H$\alpha$ velocity measurements have 
very different behaviours. The velocity curve of the narrow component 
more-or-less tracks the motion of the primary, although with substantial 
scatter. There may be an offset of roughly 0.05 
phase untits, although the scatter in the measurements 
makes determining the significance of this shift problematic. 
The half amplitudes of the velocity variations and mean velocity assuming that 
the narrow component velocities vary in sync with the primary are listed in 
Table 2. 

	The velocities obtained from the broad component show complex 
behaviour and do not track the motion of either the primary 
or secondary. However, the procedure used to obtain 
velocities from the emission line assumes a symmetric profile. 
The redward portion of H$\alpha$ emission in Figure 8 changes 
with phase, in the sense that the lower portion of the H$\alpha$ 
emission is wider at phase 0.1 than at phase 0.5, while the blue part of the 
profile is largely unchanged. The lower portions of the profile are 
thus asymmetric, with the degree of asymmetry varying with phase.
Emission from the apparently receding sub-component will bias the velocity 
measurements of the broad component at some phases to lower (i.e. 
more negative) velocities than would otherwise be expected from the 
broad component alone. The amplitude of the emission 
from the component causing the asymmetry is such that 
it has much less of an effect on the narrow component velocities, 
as those measurements were made higher up in the H$\alpha$ profile.

	\citet{broetal2021} use a combination of spectroscopic 
and interferometric observations to dissect the 
H$\alpha$ emission from $\beta$ Lyrae. They identify a jet component
that pierces the opaque disk that surrounds the secondary. 
If a jet is emerging from the disk around the secondary in BM Cas then 
this could be another possible source of H$\alpha$ emission, such as the 
red shifted component seen at phase 0.75 in Figure 8.

	The kinematics of the emission and absorption components provide 
insights into the nature of the circumsystem shell. The $\gamma$ velocity 
determined from H$\alpha$ absorption suggests that this feature forms 
in a medium that is expanding with respect to the system center of mass 
as determined from the SiII and FeII lines. If the 
shell is expanding uniformly and is azimuthally symmetric 
then wavelength shifts in the emission component from the approaching and 
receding portions of the shell should cancel out. Radial velocity variations 
in the absorption component would also not be expected. However, 
this is not the case. The variation of the overall strength of the emission 
with phase, coupled with the radial velocity variations in the absorption 
and emission components, suggest that either 
the azimuthal distribution of the shell material is asymmetric, 
and/or that the location of the ionizing radiation source varies 
with orbital phase. As was the case for H$\alpha$ absorption, 
the $\gamma$ velocity from the narrow emission 
component in Table 2 is lower than that estimated from the SiII lines, 
although we caution that the emission line system velocity relies on 
the material being in a circular orbit, which the measurements shown in 
Figure 9 suggest may not be the case.

	H$\alpha$ emission tracks the motion of Be stars in 
binary systems \citep[e.g.][]{wanetal2023}, as the emission 
originates predominantly in an accretion disk around that star. 
In the case of BM Cas the velocity curves of the 
H$\alpha$ absorption and narrow emission are in approximate phase with that of 
the primary, consistent with that star being the source 
of the majority of photons that define the H$\alpha$ profile.
If, as suggested by the light curve, the primary fills, or is close 
to filling, its Roche lobe then it is unlikely to have a circumstellar 
shell from which the emission could originate. Instead, we suggest that 
the emission comes from a circumsystem shell. If the primary 
does fill its Roche surface, then the circumsystem 
material may have been ejected through L2. This is consistent with 
the phase dependence of the amplitude of the emission. 

\section{MODELLING THE LIGHT CURVE}

	A preliminary grid of possible system elements that 
are based on the light and radial velocity curves were obtained using 
the PHOEBE \citep[][]{prsandzwi2005} modelling package. 
PHOEBE uses the Wilson-Devinny code \citep[][]{wilanddev1971,wil1979,wil1993} 
as the basis for generating models. The lack of kinematic 
information for the secondary, combined with 
the absence of a secondary eclipse and the non-orbital photometric variations 
in the light curve, introduce problematic uncertainties in the 
characterization of that star, compromising efforts to estimate reliable 
system elements. Therefore, a range of solutions in 
which the mass ratio was fixed, were examined to produce sets 
of possible elements from which general conclusions might be drawn. 
Given the loose physical constraints, no effort was made to improve upon 
the system elements found from initial comparisons with the observations 
by applying differential corrections or other optimization procedures. 
In fact, system elements that predict very different evolutionary 
states for BM Cas produce synthetic light curves that match the observations.
Still, the width of primary minimum and the variation 
of light outside of primary eclipse, some of which can be attributed to the 
tidal distortion of the primary, allow loose boundaries for the mass ratio and 
the relative size of the primary to be defined.

\subsection{Light Curves}

	We consider three data sets for the light curve 
analysis, and discuss these in turn. Only multi-cycle 
observations were considered, as they provide insights into long 
term variability that may not be related directly to the orbital motions of 
the components. The first set of observations are from the 
AAVSO APASS survey. $V$ observations of BM Cas that span a number of 
orbital cycles were recorded. While observations in $B$ were also recorded 
as part of the APASS survey, these have limited phase coverage, and 
so are not considered further. 

	A second source of photometric information are 
the observations discussed by \citet{ferandeva1997}. While 
not as extensive as the AAVSO data, these data sample a different epoch 
and so provide additional insights into possible long term variations. They 
also sample many orbital cycles, thereby allowing the suppression of 
variability that is not associated with the motions of the components.

	A third photometric dataset is that discussed 
by \citet{kaletal2009}. This compilation includes measurements made in 
U, B, V, and R. The B and V observations 
are extensive, spanning 55 orbital cycles, and 
so they are considered below. These observations overlap 
with the time interval sampled by \citet{ferandeva1997}.
The U and R observation cover a more limited time span, and so are 
not considered further given the presence of non-orbital photometric 
variations. The U and R observations will be useful for 
providing future constraints on system properties when they are combined with 
other observations in these same filters.

	The APASS, \citet{ferandeva1997}, and 
\citet{kaletal2009} light curves are shown in Figures 
10a, 10b, and 10c. The top panel in each figure shows 
$V$ magnitudes plotted according to orbital cycle, with the dotted red lines 
marking the approximate bright and faint limits of the light curve outside 
of primary minimum. Cycle-to-cycle differences in the bright and faint 
limits outside of primary minimum amount to a few hundredths of a magnitude. 
Given that the phase coverage is not uniform during each cycle then this level 
of agreement suggests that any short term variations in the light curves had 
similar amplitudes over the roughly one hundred and fifty orbital cycles 
spanned by the observations.

\begin{figure}
\figurenum{10a}
\epsscale{1.0}
\plotone{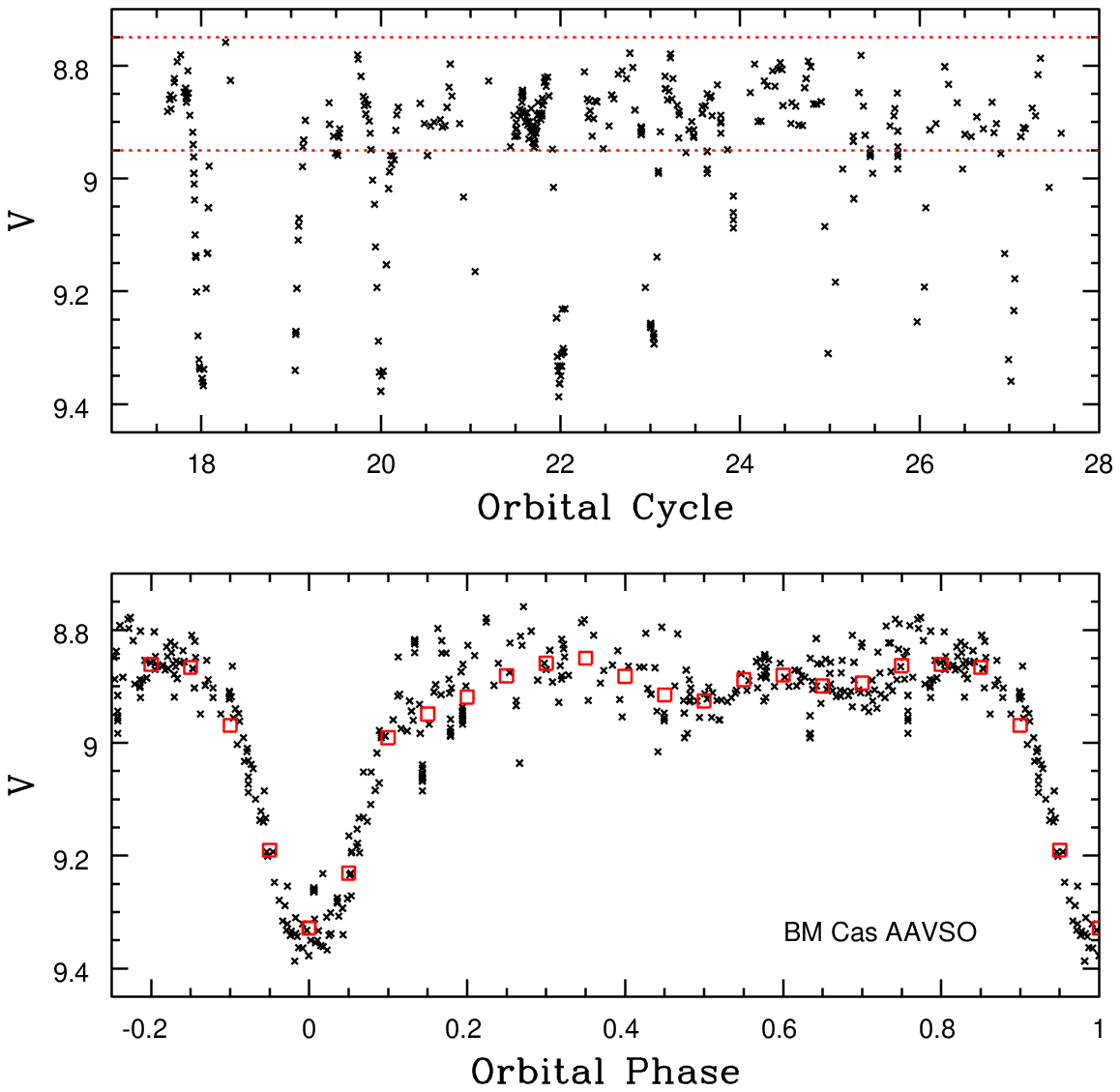}
\caption{$V$ photometry of BM Cas from the AAVSO APASS survey. 
The top panel shows observations plotted with respect 
to orbital cycle, based on the \citet{sametal2017} ephemeris, to 
examine possible long-term trends. A modest block of observations that were 
recorded near 8 orbital cycles have been excluded from the plot to 
facilitate the examination of cycle-to-cycle behaviour. The dotted 
red lines are the approximate bright and faint limits of points 
outside of primary minimum in the phased light curve. There is no evidence for 
cycle-to-cycle differences in the bright and faint limits greater than a few 
hundredths of a magnitude, as expected if the variations 
have a repeatable amplitude with a timescale that is 
markedly shorter than the system orbit, as found by 
\citet{ferandeva1997}. The phased light curve is shown in the bottom panel, 
where the red squares are normal points calculated with $\pm 0.025$ wide 
binning in orbital phase.}
\end{figure}

\begin{figure}
\figurenum{10b}
\epsscale{1.0}
\plotone{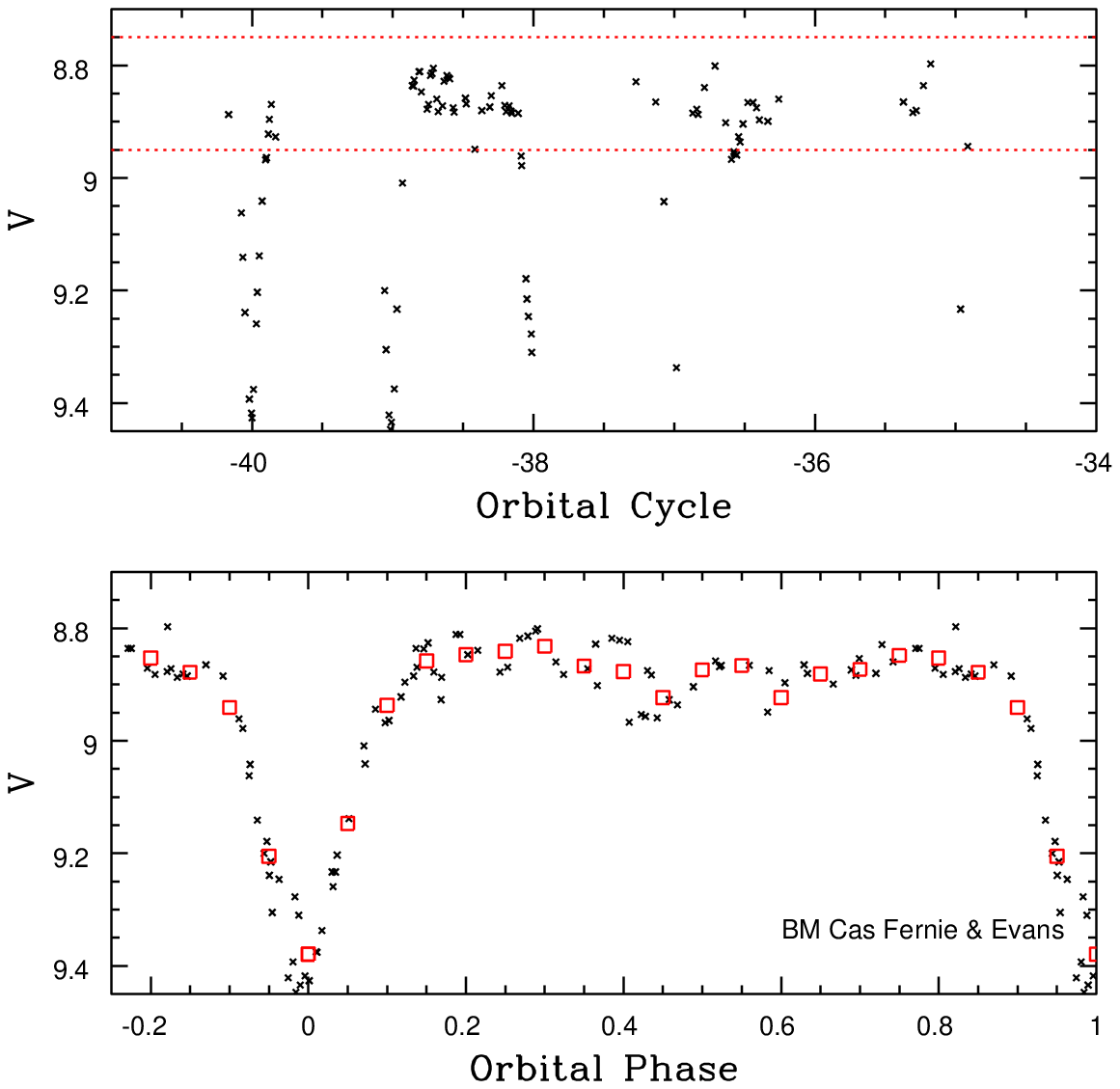}
\caption{Same as Figure 10a, but showing photometry from \citet{ferandeva1997}. 
The dashed red lines in the top panel are from Figure 10a, 
which are based on the larger AAVSO dataset.}
\end{figure}

\begin{figure}
\figurenum{10c}
\epsscale{1.0}
\plotone{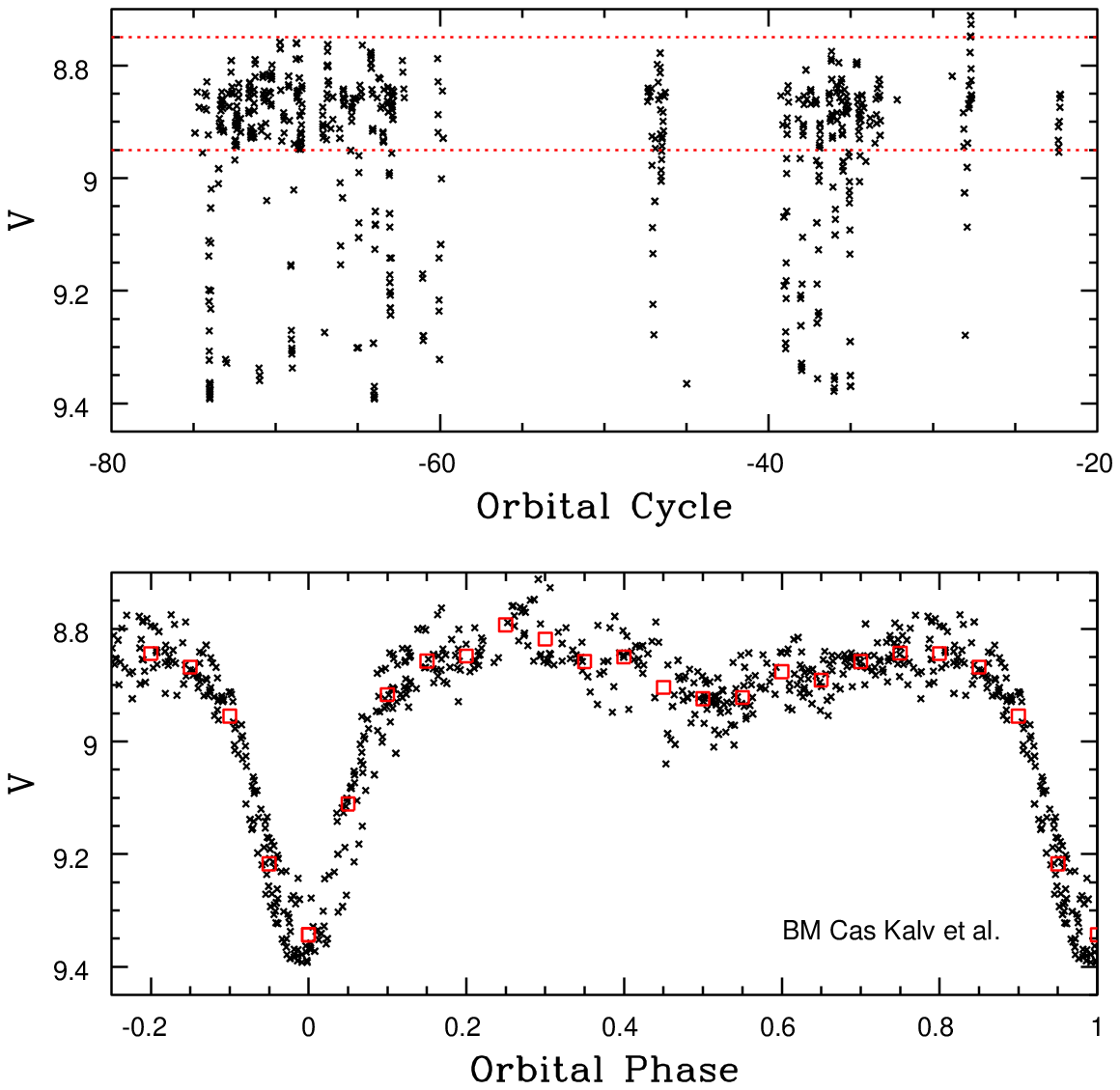}
\caption{Same as Figure 10a, but showing photometry from 
\citet{kaletal2009}. The dashed red lines in the top panel are from Figure 10a, 
and are based on the AAVSO dataset.}
\end{figure}

	Phased light curves are shown in the bottom panels of Figures 
10a, 10b, and 10c. In addition to variations that might be expected due to 
tidal distortions and the reflection effect, 
there is a dispersion of a few tenths of a magnitude
in the measurements at a given phase. The dip in the light curve near 
phase 0.5 is not due to the eclipse of the secondary, but 
is instead due to the tidal distortion of the primary. 

	Primary minimum is relatively broad, although there is 
considerable scatter near phase 0.0, adding uncertainty 
to the depth and width of this critical feature. 
There is an apparent offset in the time of primary 
minimum for some of the observations in Figure 10c, 
and \cite{kaletal2009} report 'sporatic displacements' of primary 
minimum by 0.02 phase units in some of the observations that are included 
in that figure. Persistent long-term structures 
during ingress and egress of primary minimum have been reported 
\citep[e.g.][]{shaandgap1962} although these are not present at all epochs in 
the light curves discussed here. 

	Normal points were computed in $\pm 0.025$ wide bins in orbital 
phase for each dataset, and the results are shown 
as red squares in Figures 10a, 10b, and 10c. 
The binning factor is a compromise between the desire to reduce 
the scatter in the observations while also maintaining phase resolution. 
These normal points are the basis for the light curve modelling experiments 
discussed later in this section. While the use of normal points suppresses much 
of the scatter in the light curve, clusters of points that were recorded during 
the same orbital cycle and that dominate a given bin may skew the normals. 
This appears to be the case for the AAVSO normals 
centered on phase 0.15 and 0.20, where the light curve 
is depressed in many (but not all) of the observations in this phase interval.  

	There is also a modest O'Connell Effect \citep[][]{oco1951,mil1968} in 
all three sets of observations, as the peak light level in the normal points 
between phases 0.15 - 0.40 is brighter than the peak between phases 
0.6 and 0.85 in both light curves. This could be indicative of 
spots on one of the components, presumably the 
primary. While the scatter in the unbinned light 
curve brings into question the significance of the difference in maxima, 
that it is seen in all three datasets suggests that it is real. 

	Light curves at other wavelengths would certainly be of interest, and 
observations in the IR would arguably be of great interest for understanding 
the nature of the secondary and its circumstellar environment. 
The source catalogue generated from the NEOWISE 
survey \citep[][]{maietal2014} contains a wealth 
of photometry of eclipsing binaries in the W1 and W2 filters, with 
over 400 entries for BM Cas in both filters. Unfortunately, the 
signal falls within the saturation limits of these 
filters, with the extent of saturation being greatest for the W1 observations. 
Therefore, the W1 and W2 light curves are not considered here.

\subsection{Exploring Parameter Space}

	The normal points calculated from the AAVSO, 
\citet{ferandeva1997}, and \citet{kaletal2009} 
light curves form the basis for the light curve analysis. The 
phases coming out of primary minimum in the AAVSO observations appear 
to be skewed, and it is not clear to what extent this affects primary minimum 
within that light curve. Therefore, matching the \citet{ferandeva1997} 
and \citet{kaletal2009} observations was assigned a higher priority. 
The latter also sample a large number of orbital cycles and there 
is generally good agreement between the two sets of normal points. 
Given the RUWE statistic for his system (Section 1) and the 
dispersion in the light curves introduced by non-orbital photometric 
variations, it was decided not to pursue solutions that involved third light.

	A grid of models were examined initially 
that span a range of mass ratios. These included mass ratios in excess of 
unity, as would occur if mass transfer has progressed to the point that the 
initially more massive star, which now appears as the A supergiant, is less 
massive than the gainer. Two sets of models that sample the approximate 
lower and upper limits of the mass ratio as set by the 
characteristics of primary minimum are discussed at greater 
length below. \citet{ferandeva1997} and \citet{pusetal2007} both assumed 
mass ratios less than one (i.e. the A supergiant is the more massive star). 

	The secondary has not been detected at visible and near-IR wavelengths, 
and so it is assumed to be completely obscured by a dust shell at 
visible wavelengths, as is the case in systems such as $\beta$ 
Lyrae \citep[e.g.][]{hua1963,mouetal2018,broetal2021}. Thus, the only 
source of light at visible wavelengths in these models 
is the primary. Third light was not considered given the problematic aspects 
of the light curve and the lack of information about the secondary.

	Disk parameters can be highly uncertain 
in the absence of information such as that provided by interferometric 
observations. For example, \citet{wil1979} finds that a geometrically thick 
disk at optical wavelengths is required to fit the light curve of 
$\beta$ Lyrae, whereas interferometric observations discussed by 
\citet{broetal2021} indicate a thinner, bi-modal disk geometry. 
Given these uncertainties, two geometries for the disk in BM Cas 
were considered: one in which the disk hugs the surface of the secondary 
and is in synchronous rotation with the primary, and another in which the 
disk is rotationally flattened to the point where it comes into contact with 
the critical Roche surface.

	Broad conclusions were drawn from these initial attempts to match 
model light curves with the observations. First, 
the primary and secondary must be close to filling, or 
are filling, their Roche surfaces in order to reproduce the width of 
primary minimum. Second, light curves that adopt mass ratios of a few tenths 
or less do not reproduce the properties of primary minimum. 
\citet{ferandeva1997} and \citet{pusetal2007} reached similar conclusions 
regarding the size of the primary and the system mass ratio. Finally, mass 
ratios in excess of one (i.e. the primary has a mass that is lower 
than the secondary) also match the light curves. Mass ratios much larger 
than 2 are ruled out, as they can not fit the width of primary 
minimum, and do not provide the degree of tidal distortion of the primary 
that is required to fit the depth of the light curve at phase 0.5. 
An independent estimate of the mass ratio that is based on the 
rotational velocity of the primary based on the 
procedure described by \citet{har1990} is presented 
in the Appendix. While there are considerable uncertainties, this procedure 
predicts a mass ratio that is intermediate between 
the values considered here.

\subsection{Specific Models}

	We consider two specific models that assume very different 
evolutionary states for the system. In both cases the secondary 
is assigned an effective temperature of 1000K, and experiments showed 
that adopting other effective temperatures that are less 
than $\sim 1500$K had negligible effects on model light curves in $V$. 
We caution that the system elements that result from these 
efforts should not be considered to be definitive, as other elements are 
possible. Rather, this modelling effort is simply intended to demonstrate 
that a wide family of parameters can match the existing observations, and 
that models in which the system has undergone mass transfer are possible.

	Linear limb-darkening (x) and gravity darkening ($\beta$) 
coefficients were based on the solar metallicity models described 
by \citet{claandblo2011}. For the primary, x=0.5 and $\beta = 0.6$ 
were adopted. Unfortunately, there is considerable uncertainty in the 
values for the secondary, as the models do not go fainter than 
an effective temperature of 3500K. Still, there 
is a trend of x increasing as T$_{eff}$ drops for low gravity models with 
T$_{eff} \leq 5000$K, and so x=1 was adopted for the secondary. 
The solar metallicity \citet{claandblo2011} models for low surface 
gravities have $\beta \sim 0.6$ when T$_{eff} \leq 5000$K, and so this 
was also adopted for the secondary. Finally, a bolometric 
albedo of 0.9 was adopted for the primary based on the models discussed 
by \citet{cla2001}, while for the secondary 0.5 was adopted based 
on the expectations for a convective star \citep{ruc1969}.

	The radial velocity variations in the SiII lines were also 
modelled with the light curves to compute absolute dimensions. All models 
produce a smoothly varying sine curve, replicating the trend seen in the data, 
and so emphasis was placed on reproducing the overall amplitude of the 
velocity curve. Therefore, we do not show a comparison between the 
observed and modelled velocity curves. The radial 
velocity variations were best matched with a 
system velocity of -58 km/sec, which falls within the uncertainty of the 
system velocity in Table 2 that was measured from the SiII lines.

	Model light curves are compared with normal points 
in Figures 11 (q=0.5) and 12 (q=2.0). The top panels assume a secondary 
that is in synchronous rotation with the primary, while the bottom panels 
show light curves in which the secondary is flattened by asyncronous rotation 
to produce a disk-like structure -- these are referred to as 
the 'flattened disk' models.

\begin{figure}
\figurenum{11}
\epsscale{1.0}
\plotone{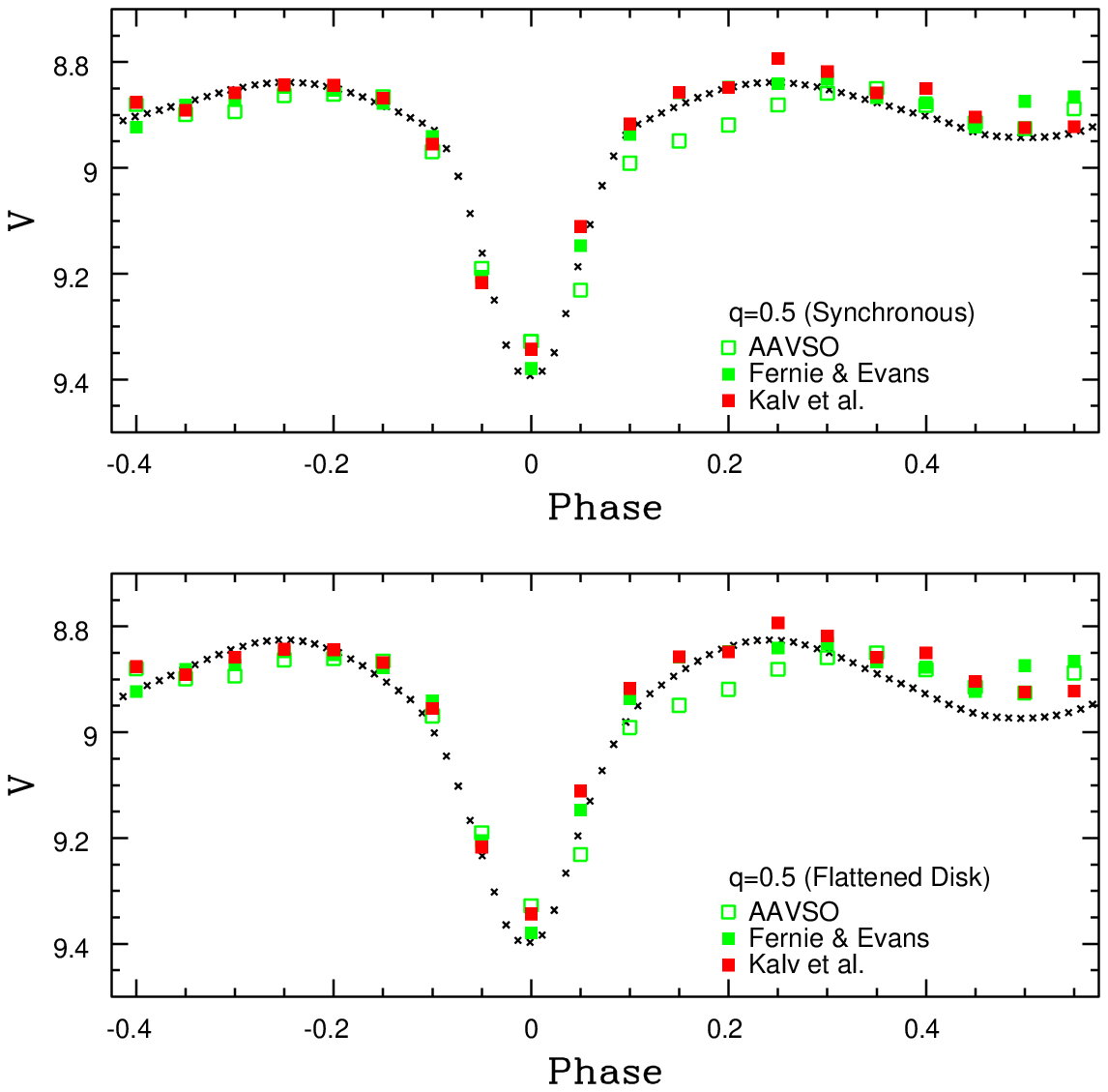}
\caption{Comparing model light curves with an assumed mass ratio of 0.5 
to normal points from the AAVSO (open green squares), \cite{ferandeva1997} 
(filled green squares) and \citet{kaletal2009} (filled red squares) light 
curves. This mass ratio is consistent with an 
evolutionary model in which mass transfer has eiher not yet 
started, or is in its very early stages. The model light curve in the top panel 
assumes that the secondary is a very cool object in synchronous rotation 
with the system, while that in the lower panel assumes that the secondary is 
rotationally flattened disk, with a length along 
the system axis that touches the critical Roche surface 
of that star. The effects of long-term variablity in the AAVSO light curve 
are clearly evident, especially near primary minimum and between phases 
0.1 and 0.3. The models provide a better match to the depth of primary minimum 
in the \citet{ferandeva1997} and \cite{kaletal2009} observations, although 
there is considerable dispersion in the unbinned measurements at this phase 
in all three datasets. The AAVSO and \citet{kaletal2009} observations near 
phase 0.5 provide a better match to the model 
than the \citet{ferandeva1997} observations. That 
the models do not match the normal points near phase -0.25 suggests 
that the surface of the primary may be spotted.}
\end{figure}

\begin{figure}
\figurenum{12}
\epsscale{1.0}
\plotone{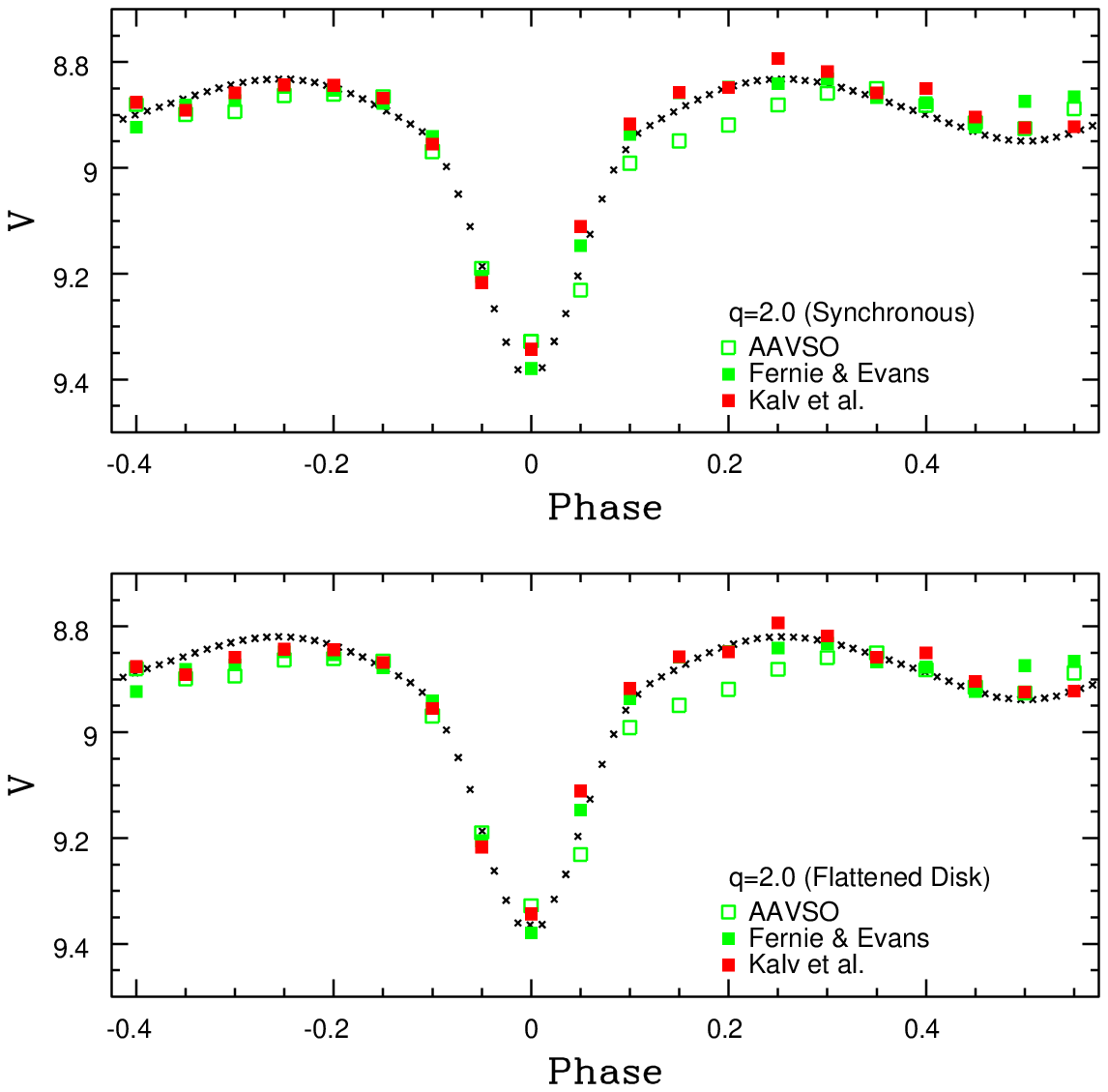}
\caption{Same as Figure 11, but showing model light curves for a 
mass ratio of 2.0. This mass ratio is consistent with an evolutionary state 
in which significant mass transfer has taken place, such that the initially 
more massive primary is now the less massive 
star. While the agreement with the observations is 
far from perfect, the models with q=2.0 are better able to reproduce the 
width and depth of primary minimum as defined by the \citet{ferandeva1997} 
and \citet{kaletal2009}} observations than the q=0.5 models.
\end{figure}

	The B--V colors from the \citet{kaletal2009} photometry 
are compared with those predicted from the models in Figure 13. 
There is only a few hundredths of a magnitude variation
in B--V with phase. The overall difference between the observed and 
predicted colors suggest E(B--V) = 1.01 magnitudes, in agreement with that 
estimated by \citet{ferandeva1997}, and the modelled color curves have been 
shifted by this amount. There is a significant difference between the 
observed and predicted color near primary minimum. With the exception of 
the system inclination, the B light curve can be reproduced with 
the same elements as those found for V. The B light curve requires an 
angle of inclination that is a few tenths of a degree larger than that for V. 
As the depths of primary minimum in V and B can not be fit with the 
same orbital inclination, we suspect that there might be a diffuse dust 
component around the secondary that has not been included in the models.

\begin{figure}
\figurenum{13}
\epsscale{1.0}
\plotone{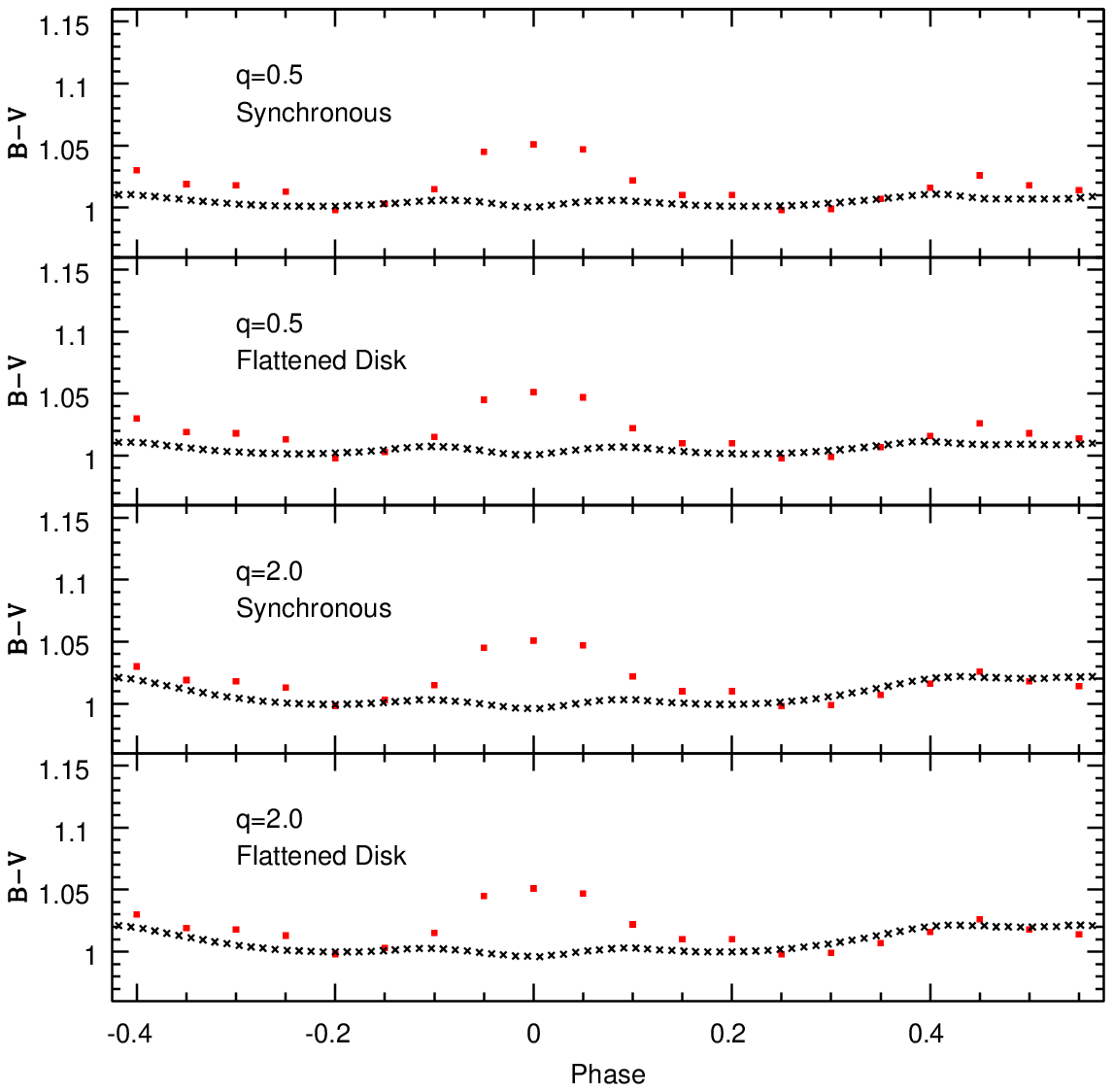}
\caption{Comparing observed and predicted B--V colors. 
The open squares are B--V colors computed from the 
\citet{kaletal2009} normal points. The model colors have been shifted 
by 1.01 mag to account for line of sight reddening. The models 
do not reproduce the reddened color near primary minimum, 
and we suspect that this is due to a diffuse dust halo around the 
secondary that is not included in the models (see text). The q=2.0 models 
reproduce the observed color outside of primary minimum better than the 
q=0.5 models.}
\end{figure}

	The \citet{ferandeva1997} and \citet{kaletal2009} light 
curves provide a better match to the models 
than the AAVSO observations. The models do not match the depth and shoulders 
of primary minimum in the AAVSO light curve, although the 
models better match the AAVSO and \citet{kaletal2009} normals near phase 0.5. 
The difference in brightness at phases -0.25 and 0.25 
also suggest that the surface of the primary 
may be spotted. However, for this initial examination of the 
light curves we have opted not to include spots in the light curve models. 
The q=0.5 and q=2.0 models are discussed at length in the following 
subsections.

	System elements for the models are listed in Table 3, where 
A is the semi-major axis in solar radii. Experimentation suggests 
that the orbital inclination has an uncertainty of 
$\pm 1 - 2$ degrees, whereas the masses are likely uncertain by 
$\pm 2 - 3$M$_{\odot}$ for q=0.5, and $\pm 0.5 - 1$M$_{\odot}$ 
for q=2. The uncertainties in the stellar radii are 
likely on the order of a few hundreths of the orbital separation. 
These uncertainties are estimates only, and apply only for the assumed 
mass ratio; they do not reflect the true uncertainties in the actual 
system elements.
 
\begin{deluxetable}{cllll}
\tablecaption{Preliminary Light Curve Elements}
\tablehead{Element & q=0.5 & q=0.5 & q=2.0 & q=2.0 \\ & Synchrous & Flattened Disk\tablenotemark{a} & Syncronous & Flattened Disk\tablenotemark{a}\\}
\startdata
T1$_{eff}$ (K) & 8500 & 8500 & 8500 & 8500 \\
$\Omega_1$ & 3.1 & 2.9 & 5.45 & 5.45 \\
r1$_{pole}$ & 0.38 & 0.41 & 0.28 & 0.28 \\
r1$_{side}$ & 0.40 & 0.43 & 0.29 & 0.29 \\
r1$_{point}$ & 0.44 & 0.53 & 0.34 & 0.34 \\
r1$_{back}$ & 0.41 & 0.46 & 0.32 & 0.32 \\
M$_1$ (M$_{\odot}$) & 68 & 68 & 5.6 & 5.6 \\
R$_1$ (R$_{\odot}$) & 265 & 265 & 107 & 107 \\
L1$_{V}$ & 1.0 & 1.0 & 1.0 & 1.0 \\
 & & & & \\
T2$_{eff}$ (K) & 1000 & 1000 & 1000 & 1000 \\
$\Omega_2$ & 3.5 & 3.5 & 5.45 & 5.4 \\
r2$_{pole}$ & 0.22 & 0.22 & 0.40 & 0.40 \\
r2$_{side}$ & 0.22 & 0.26 & 0.42 & 0.43 \\
r2$_{point}$ & 0.23 & 0.30 & 0.48 & 0.52 \\
r2$_{back}$ & 0.23 & 0.28 & 0.44 & 0.46 \\
M$_2$ (M$_{\odot}$) & 34 & 34 & 11 & 11 \\
R$_2$ (R$_{\odot}$) & 150 & 150 & 152 & 157 \\
L2$_V$\tablenotemark{b} & 0.0 & 0.0 & 0.0 & 0.0 \\
 & & & & \\
i & 82 & 80 & 63.5 & 63.0 \\
A (R$_{\odot}$) & 665 & 665 & 365 & 365 \\
\enddata
\tablenotetext{a}{Assumes rotation that maximizes the extent of the 
secondary without contacting the critical Roche surface.}
\tablenotetext{b}{The secondary is assumed to be completely obscured 
at visible wavelengths.}
\end{deluxetable}

\subsubsection{q=0.5}

	If the mass ratio is 0.5 then the primary is more massive than the 
secondary; mass exchange has not started, or is at a very early stage. Changes 
in the orbital period are a potential test for the latter case. In particular, 
if mass exchange has just started then large changes in the period are 
to be expected, as the rate of mass transfer is expected to be very high as the 
separation between the components drops with time \citep[e.g.][]{nelandegg2001}.
However the non-orbital photometric variations in the BM Cas light curve, 
coupled with the long period, complicate efforts to find reliable 
times of minima.

	The \citet{ferandeva1997} observations more-or-less match the 
predicted depth of primary minimum if q=0.5, whereas the \citet{kaletal2009} 
and AAVSO observations have a shallower primary minimum. The AAVSO 
observations do not match the q=0.5 models during primary minimum egress. 
The \citet{kaletal2009} and AAVSO observations better match 
the light curve near phase 0.5, where the dip in the light curve is due to 
the tidal distortion of the primary. The flattened disk model is better 
able to match the width of primary minimum. However, the model in which the 
disk is less flattened produces a better match to the 
light curve near phase 0.5. As for B--V, both q=0.5 models 
in Figure 13 underestimate the color dispersion outside of primary minimum.

	The two disk geometries produce large differences in the 
shape of the secondary. Adopting the ratio of the 
pole and point radii as a measure of distortion, 
then the secondary in the synchronous rotation model has an eccentricity 
of 0.06, whereas the eccentricity is 0.27 with 
the flattened disk model. The primary has a larger radius with the 
flattened disk model, and hence is more distorted; this is reflected in the 
shape of the model light curve outside of primary minimum. 
An interesting aspect of the q=0.5 model is that an orbital 
inclination of 80 -- 82 degrees is needed to match the depth and width 
of primary minimum. This affects the estimated component 
masses, as previous studies predicted a smaller angle of inclination. 

	The complications in matching the observations aside, there are 
problems with the q=0.5 models. The components are very massive, with 
M$_1 = 68$M$_{\odot}$ and M$_2 = 34$M$_{\odot}$. While such masses are not 
without precedent, the primary would then be among the most massive stars 
in the Galaxy. However, there is no evidence that BM Cas is in a star-forming 
region, where stars with such high masses would be expected. The mass of the 
secondary is also such that a wind would be present, and such an outflow 
may pierce or disrupt a disk.

	The dark nature of the secondary in the 
visible creates another problem with this mass ratio, 
as mass transfer has either not started or has just 
started. In the former case then the origins of the disk are not obvious. As 
for the latter case, while systems believed to be in a similar early phase of 
mass transfer have an accretion disk around the secondary, the disk is usually 
illuminated by the mass stream interacting with the disk. The disk is then a 
source of prominent emission in the UV, and this is not 
seen in the spectra of BM Cas \citep[][]{ferandeva1997}. Finally, while not 
a fatal problem, a circular orbit for such a young, long period system 
is perhaps unexpected.

\subsubsection{q=2.0}

	Model light curves that assume a mass ratio of 2 are compared with 
the observations in Figure 12. The A supergiant is the less massive star in 
these models, and the system is thus highly evolved, with mass transfer having 
depleted the primary to the point where it is no longer the more massive star.
The overall system mass is also much lower 
than if q=0.5, with M$_1 = 5.6$M$_{\odot}$ and M$_2 = 11.2$M$_{\odot}$. 
These masses are indicative of intermediate initial masses and are 
more in keeping with the environment around BM Cas. 

	Primary minimum in the model light curves with q=2.0 has a broad bottom 
due to the large size of the secondary when compared with the primary. 
This is consistent with the behaviour of the lower portions 
of primary minimum in the \citet{kaletal2009} and AAVSO observations in 
Figures 10a and 10c, but not with the \citet{ferandeva1997} light 
curve in Figure 10b. Other aspects of the models also give mixed 
results when compared with the observations. The models 
do not match the depth of primary minimum in the 
\citet{kaletal2009} and AAVSO observations, although the overall 
width of primary minimum is well matched by both q=2.0 models. 
While the differences are modest, the q=2.0 models are a 
better match to the \citet{kaletal2009} and \citet{ferandeva1997} 
light curves near phases 0.9 and 0.1 (ingress and egress of primary 
eclipse) than the q=0.5 models. As for B--V, the q=2.0 models are 
better able to match the color curve outside of primary minimum.

	The dimensions of the secondary differ between the two disk models. 
However, differences in the the degree of flattening of the secondary are 
smaller if q=2.0 than if q=0.5, with an eccentricity of 0.16 for 
the secondary in the synchronous model, and 0.23 for the flattened disk model.
These dimensions reflect the need for the secondary to be much larger 
than the primary in order to reproduce the width of primary minimum.

	Many of the problems discussed earlier with the q=0.5 model are 
relieved if q=2.0. The component masses are more in line with what might 
be expected for stars that are outside areas of active star formation. 
That mass transfer has occured also provides a natural explanation for 
a disk around the secondary, in that an opaque shell could be the remnant of an 
accretion disk that is no longer excited by a mass stream from the primary.

\section{DISCUSSION \& CONCLUSIONS}

	Spectra that sample wavelengths near H$\alpha$ 
with a spectral resolution of $\sim 17000$ have 
been combined with archival data to explore the properties 
of the 198 day period eclipsing binary BM Cas, with a goal of 
gaining insights into the evolutionary state of the system. 
Despite the importance of such long period systems, BM Cas has only been the 
subject of a handful of studies, likely due to its period.
BM Cas is of particular interest given that 
the system mass function hints that the components may have 
large initial masses, although this depends on the mass ratio. The long 
period also suggests that mass transfer may have occured (or will occur) 
when one of the components was (is) in a more highly evolved state 
than in shorter period systems. This will affect 
the reaction of the components to mass transfer \citep[e.g.][]{ibeandliv1993} 
and the final nucleosynthetic yield \citep[e.g.][]{faretal2023}, 
among other factors. 

\subsection{System Components and Elements}

	The DAO spectra track kinematically distinct system components. Tracers 
that are attributed to photospheric, circumstellar, and circumsystem 
environments have been detected. Spectroscopic signatures of the secondary or 
the surrounding disk were not identified, and BM Cas thus remains an SB1. 

\subsubsection{The Velocity Curve of the Primary}

	The velocities that were obtained from the cores of the SiII 
and FeII lines differ. \citet{pop1977} also found a dichotomy when comparing 
velocities measured from MgII 4481 with those obtained from other lines at 
blue wavelengths. We have shown that the differences 
between the velocities measured from the core of 
FeII 6456 and those obtained from the SiII lines is due to 
a deep and narrow shell absorption feature in the former. Shell components 
are seen in other FeII lines. 

	A shell component is not obvious in the SiII lines, although 
sub-structuring in the SiII profiles hint that such a feature may be present. 
A component in the FeII lines that follows the motions of the 
SiII lines is detected but it is not well resolved with our 
spectra. The stable kinematic properties of the FeII lines suggest that the 
structure of the shell in the region where FeII absorption 
occurs did not change measurably over the five 
orbital cycles covered by the DAO spectra.

	The velocities measured from the SiII lines likely track the 
orbit of the primary, and the full amplitude of the SiII velocity 
curve is $116.6 \pm 2.8$ km/sec. The SiII lines yield 
velocities that approximately match those obtained 
from MgII 4481 by \citet{pop1977}, although there are departures of up to 
$\sim 10$ km/sec from the \citet{pop1977} measurements. The behaviour of the 
MgII 4481 line in the spectra of intermediate A type stars may be 
susceptible to contamination from the FeII and TiII blend at 4471\AA\ 
\citep[][]{graandgar1989}, and this might be 
a factor in the differences between the SiII and MgII 
velocity curves. Cycle-to-cycle variations in the SiII velocities 
also contribute noise to the radial velocity curve, and we suspect these are 
the result of surface activity on the primary.

\subsubsection{The Circumsystem Shell}

	The velocities obtained from H$\alpha$ emission depend on 
location in the line profile, reflecting the composite nature of the 
emission. Velocities measured near the peak of the profile are indicative of 
motion that is more-or-less synchronized with that of the 
primary, as expected if the source of the photons powering 
the emission is related to the primary. 
In contrast, the velocities measured near the base of H$\alpha$ 
do not track the motions of the primary, likely due to contamination 
in that part of the profile by at least one other H$\alpha$ component that 
is weaker than the emission from the shell.

	Given the kinematics of the H$\alpha$ emission measured near the
top of the profile and the close proxity of the A supergiant to its 
Roche lobe, then it is likely that this emission orignates in a 
circumsystem shell, as opposed to a circumstellar shell around 
the supergiant. Moreover, that the mean velocities of 
H$\alpha$ absorption and emission agree, and are higher (i.e. less 
negative) than the systemic velocity obtained from SiII absorption 
lines suggests that the emitting region is expanding. 
\citet{pop1977} also finds that the majority of the lines in the blue part of 
the spectrum, which we attribute to the shell, have a higher system velocity 
than that found from MgII 4481, although the statistical significance is low.

	There is a $\sim 3$ km/sec velocity variation in the 
H$\alpha$ absorption component that is in phase with the velocity curve 
of the primary. This synchronization is not unexpected if the primary, or 
an object associated with it, is the likely source of the absorbing 
photons. Still, if the shell were in a stable circular orbit 
then there should be no velocity variation with orbital phase, 
as the shell motions would always be tangential to the 
line of sight. That a velocity variation is seen suggests that the 
disk is not in a circular orbit.

	The strength of H$\alpha$ emission as gauged from the V and R 
measurements peaks near phase 0.5, when the primary is in front of 
the secondary. \citet{ferandeva1997} find tentative evidence 
for an increase in UV flux at a phase just 
past secondary minimum, as might be expected if there was a source of UV 
light on the side of the primary that is not facing the secondary. However, 
they caution that the S/N ratio is low and that the results are susceptible 
to variations in the signal that are not due to the orbital motions 
of the components. Phase 0.5 is where the L2 point 
is directed towards the viewer, and a hot spot at this point 
could enhance the strength of H$\alpha$ emission.

	A second spike in H$\alpha$ emission 
strength occurs at primary minimum, when the secondary 
is in front of the primary. \citet{ferandeva1997} find that 
the UV flux also increases at this phase, which is counter 
to what might naively be expected given that light from the hotter component 
is eclipsed. The increase in UV signal then hints at a source of UV 
light such as a hot spot on the surface of the primary 
that is facing the secondary, or an emitting region along the 
line that connects the two components, as suggested by \citep[][]{pusetal2007}. 

\subsection{Non-orbital Variablity}

	Variability in the BM Cas light curve that appears not to be 
directly related to the orbital motions of the components is 
seen over a range of wavelengths with a 
quasi-periodicity that is substantially less than the 
orbital period of the system \citep[][]{ferandeva1997}.
While BM Cas was initially thought to contain a Cepheid 
\citep[e.g.][]{thi1956}, \citet{ferandeva1997} 
conclude that Cepheid variability is not consistent 
with the characteristics of the photometric variations. In fact, 
complicated photometric variations are common in the light curves 
of interacting systems \citep[e.g.][]{fre1957}, 
and these have been attributed to accretion 
activity \citep[e.g.][]{desetal2015}. These variations 
tend to be non-periodic, and in many cases the 
source is thought to be a hot spot on the accretion disk around the 
receiving star. However, the variations seen in BM Cas 
are quasi-periodic with a time scale that is very short when compared with 
the orbital period, which is typically not the case in W Ser systems. 

	The amplitude, periodicity, wavelength dependence and erratic 
behaviour of the non-orbital variations are reminiscent of those seen 
in $\alpha$ Cyg variables. \citet{pop1977} finds that the primary of 
BM Cas has a spectral type that is similar to that of the $\alpha$ Cyg variable 
HR825 (V480 Per). If this is the case then the photometric variations in 
the light curve that are not due to the orbital motions of the components 
may provide insights into the internal structure and 
evolutionary state of the primary.

\subsection{Light Curve Modelling and the Evolutionary Status of BM Cas}

	A range of possible system elements that can be estimated from the 
light curves have been explored for different fixed mass ratios. Model 
light curves with mass ratios between 0.5 and 2.0 match the width of 
primary minimum and the properties of the light curve near phase 0.5. These 
mass ratios predict very different system properties and evolutionary states.

	If the mass ratio is 0.5 (i.e. the primary is twice as massive as 
the secondary), then the primary has a mass of 
68 M$_{\odot}$, making it an exceptionally massive star. 
In contrast, if the mass ratio is 2.0 (i.e. the secondary is two 
times more massive than the primary), then the mass of the primary drops 
to 6 M$_{\odot}$ and the overall system mass is more in line with that of 
many other Algol systems. Of course, there is a continuum of possible 
solutions between (and probably slightly outside of) the 
range of mass ratios considered here.

	Determining reliable properties of the component stars 
from the existing photometry is problematic given the various sources 
of uncertainty, and this is demonstrated in some of the light curve elements, 
such as the orbital inclination. The most significant sources of 
uncertainty are the nature of the secondary and its circumstellar environment. 
Photometric variations that are not due to the orbital motions and 
the tidal distortions of the components are a compounding source 
of uncertainty when modelling the light curve, and these have a substantial 
impact on the depth and shape of primary minimum. While primary minimum is 
of greatest importance for assessing light curve models, these variations 
also affect the light curve at other phases, where constraints on the tidal 
distortion of the primary can be gleaned.
The presence of circumstellar and circumsystem 
material is another source of uncertainty, as this has the potential to skew 
temperature estimates and component velocities. 

	Despite the difficulties in extracting firm system elements from the 
light curve, indirect clues into the nature of BM Cas 
can be gleaned from its environment. In particular, the 
extent of ISM emission along the line of sight is typical of 
its surroundings, suggesting that BM Cas is not in a dense molecular 
environment. The proper motions of BM Cas indicate that it is 
not a runaway star, such as might be expected if 
it had been ejected from a star-forming region; in fact, the proper motions 
are typical of other stars in its vicinity that were 
selected based solely on parallax and location on the sky. 
The projected separation from stars with $G < 18$ that have 
parallaxes and proper motions that are similar to those of BM Cas is $\sim 
1 - 2$ parsec, further indicating that it is not in a crowded 
star-forming environment. While there are 
faint candidate PMS objects within a projected separation of $\sim 6$ 
parsecs of BM Cas, that there is considerable depth in the extraction volume 
along the line of sight makes a connection with BM Cas uncertain. 
When considered in concert, these indicators are consistent with an 
an age in excess of a few tens of Myr; the primary and secondary then have 
intermediate, as opposed to high, masses. 

	The most distinct attribute of BM Cas when compared with its 
surroundings is that it is far brighter than any of its neighbors. This is a 
property that is common among intermediate mass systems in the field that are 
experiencing or have experienced mass transfer \citep[e.g.][]{dav2023}. 
Arguably the most convincing sign that there has been mass transfer 
is the evidence for an optically thick disk around the secondary
that is likely the remnant of an accretion disk that is no longer 
illuminated by a mass stream from the primary. 

	There are other interacting systems that have a heavily obscured 
secondary, and two examples are $\beta$ Lyrae and 
$\epsilon$ Aurigae \citep[][]{kloetal2010,mouetal2018}. 
$\beta$ Lyrae (P = 12.9 days) and $\epsilon$ Aur (P = 9896 days) have 
orbital periods that bracket those of BM Cas, and interacting systems with 
dark disks that have even longer periods than that of $\epsilon$ Aur 
have been found \citep[e.g.][]{rodetal2016}. The broad range of periods in 
which obscuring disks are seen suggests that these 
structures can form in systems in which mass transfer occurs when the primary 
is in a wide range of evolutionary states.

	A circumsystem shell is another likely artifact 
of mass transfer. Evidence for a circumsystem shell is seen 
in the general character of H$\alpha$ and in the 
narrow absorption features in the FeII lines. 
A shell may form as a result of non-conservative 
mass transfer, and shells are found around many other interacting systems. 
Finally, while not an ironclad indicator of interactions, BM Cas has a 
circular orbit, despite having a relatively long period. Given 
that the frequency of eccentric orbits increases with 
period \citep[e.g.][]{lietal2023} then a circular orbit is consistent with 
mass transfer having taken place via angular momentum exchange. 

	We conclude that BM Cas is likely an evolved binary 
system in which mass transfer has occured, but 
ended when the primary detached from its Roche surface. The 
primary has lost mass, and is now the less massive component. The secondary 
has gained mass, and is currently embedded in an opaque disk, similar to 
that around the secondary in $\beta$ Lyrae.

\subsection{Future Work}

	The only observational signatures of the secondary at visible 
wavelengths are its gravitational influence on the primary and the attenuation 
of light during primary minimum. Therefore, 
the greatest contributions to understanding BM Cas will 
almost certainly come from observations that cast light on the nature of the 
secondary and accompanying circumstellar material. Additional observations 
in the IR should be rewarding, as suggested by \citet{pusetal2007}. 
For example, IR light curves have provided important checks on models of 
the circumstellar material in the $\beta$ Lyrae system 
\citep[e.g.][]{zeietal1982}. Unfortunately, while there are multiple 
W1 and W2 measurements of BM Cas in the NEOWISE archive, 
saturation is an issue. Observations in W3 and W4 recorded as part 
of the ALLWISE program sample only three points 
in the light curve, and these are outside of primary minimum. 

	That the primary has become detached from its Roche lobe, or 
is in contact with it, could be confirmed from more detailed 
modelling of the light curve. In addition to light curves in the 
IR, the analysis of additional light curves at visible 
wavelengths are also essential to establish the 
characteristics of the long-term photometric variations that cause 
the large differences between the light curves, and thereby allow the 
photometric components solely due to the motions of the components to be 
isolated. It is worth noting that, with 
the exception of the depth of primary minimum, 
the \citet{kaletal2009} and \cite{ferandeva1997} light curves are 
similar, even though the former covers a much longer time span than the latter. 
This suggests that the light curve outside of eclipse was relatively stable 
when the \cite{kaletal2009} data were obtained. Still, the AAVSO 
light curve, which was recorded a few tens of orbital cycles later, is very 
different from the others. This indicates that observations covering at 
least many hundreds of orbital cycles will likely be necessary to characterize 
any long term behavior, assuming that the long term 
variations are periodic, rather than stochastic. 

	The chemical abundances of the primary will provide a fossil 
record of the past evolution of that star. If, as we 
suspect, BM Cas is an evolved Algol system then the primary was 
the source of the mass flow, and so its chemical abundances might show 
mixtures that differ from those of A supergiants that are not 
in binary systems. A potential complicating factor is that the chemical 
mixtures of stars in unresolved binary systems may be inherently 
different from those in isolated stars \citep[e.g.][]{pavetal2023}. 
One way to mitigate against this would be to consider 
A supergiants in binaries that have not undergone mass transfer as a 
possible comparison sample for BM Cas. A measurement of the $^{12}C/^{13}C$ 
ratio in the circumsystem shell, that presumably is composed of material 
ejected from the primary, might also provide information about mass 
transfer \citep[][]{faretal2023}.

	Finally, observations in the UV may also provide additional 
insights into the evolutionary status of the system. 
UV emission is one signature of mass transfer, 
and such emission has been detected from systems that 
contain a heavily obscured secondary, such as $\beta$ 
Lyrae \citep[e.g.][]{plaandkoc1978}. While \citet{ferandeva1997} obtained 
UV spectra of BM Cas, these have a low S/N, and so can be improved upon. 
The behaviour of H$\alpha$ emission suggests that 
enhanced UV emission might be expected near phase 0.5. A UV spectrum near 
phase 0.6 was obtained by \citet{ferandeva1997}. 

\acknowledgements{It is a pleasure to thank the anonymous referee for 
comments that greatly improved the manuscript. Sincere 
thanks are also extended to David Bohlender, Dmitry Monin, and Deb 
Lockhurst for their tireless efforts working with the DAO telescopes. 
This paper is based on observations obtained at the Dominion Astrophysical 
Observatory, NRC Herzberg, Programs in Astronomy 
and Astrophysics, National Research Council of Canada. The research 
presented in this paper has also used data from the Canadian Galactic 
Plane Survey, a Canadian project with international partners, supported 
by the Natural Sciences and Engineering Research Council. This work 
also has made use of the WISE image (https://doi.org/10.26131/irsa151) 
and NASA Extragalactic Database services
(https://doi.org/10.26132/ned1), at the NASA/IPAC Infrared Science Archive 
which is funded by the National Aeronautics and Space Administration and
operated by the California Institute of Technology. Finally this study has 
also used data from the European Space Agency (ESA) mission
{\it Gaia} (\url{https://www.cosmos.esa.int/gaia}), processed by the {\it Gaia}
Data Processing and Analysis Consortium (DPAC,
\url{https://www.cosmos.esa.int/web/gaia/dpac/consortium}). Funding for the DPAC
has been provided by national institutions, in particular the institutions
participating in the {\it Gaia} Multilateral Agreement.} 

\appendix

\section{The Rotation Velocity of the A Star and the Mass Ratio}

	The mass ratio is a key quantity for understanding 
the properties of BM Cas. \citet{har1990} describes an iterative technique to 
estimate the mass ratio of a system if one of the stars is assumed to 
fill its Roche lobe and is in tidally-locked rotation. The core observational 
element is a reliable measurement of {\it vsini} for that star. 
The DAO spectra are not ideal for estimating {\it vsini} given that 
the SiII lines are unresolved doublets, while the FeII lines are contaminated 
by shell absorption. However, \citet{glaetal2008} have estimated the 
projected rotational velocity of the A supergiant 
from the MgII 4481 line, which appears to be free of  
shell absorption features, and find $V_{rot}$sini = 57 km/sec. 
An uncertainty was not given.

	Applying the procedure described 
by \cite{har1990} reveals that the estimated size 
of the A supergiant is smaller than predicted if q=0.5, while this star 
is larger than predicted if q=2.0. Approximate agreement between the estimated 
and predicted size of the A supergiant is found for mass ratios near 
unity. We caution that this procedure is not free of uncertainties, as it is 
not clear if the primary (1) is in contact with its Roche surface and (2) 
is in synchronous rotation. Still, it is encouraging that a 
mass ratio that falls between the limits set from the analysis of the 
light curves is predicted.


\parindent=0.0cm

\begin{thebibliography}

\bibitem[A. Armeni et al. (2022)]{armandsho2022}
Armeni, A., \& Shore, S. N. 2022, A\&A, 664, A103

\bibitem[M. S. Bessell \& J. M. Brett (1988)]{besandbre1988}
Bessell, M. S., \& Brett, J. M. 1988, PASP, 100, 1134

\bibitem[W. Bidelman (1982)]{bid1982}
Bidelman, W. 1982, IBVS, No. 2112

\bibitem[W. R. Brown et al. (2020)]{broetal2020}
Brown, W. R., Kilic, M., Kosakowski, A., et al. 2020, ApJ, 889, 49

\bibitem[M. Broz et al. (2021)]{broetal2021}
Broz, M., Mourard, D., Budaj, J. et al. 2021, A\&A, 645, A51

\bibitem[K. B. Burdge et al. (2020)]{buretal2020}
Burdge, K. B., Prince, T. A., Fuller, J., et al. 2020, ApJ, 905, 32

\bibitem[J. A. Cardelli et al. (1989)]{caretal1989}
Cardelli, J. A., Clayton, G. C., \& Mathis, J. S. 1989, ApJ, 345, 245

\bibitem[J. M. Carpenter (2001)]{car2001}
Carpenter, J. M. 2001, AJ, 121, 2851

\bibitem[A. Claret \& S. Bloemen (2011)]{claandblo2011}
Claret, A., \& Bloemen, S. 2011, A\&A, 529, A75

\bibitem[A. Claret (2001)]{cla2001}
Claret, A. 2001, MNRAS, 327, 989

\bibitem[T. J. Davidge (2022)]{dav2022}
Davidge, T. J. 2022, AJ, 164, 149

\bibitem[T. J. Davidge (2023)]{dav2023}
Davidge, T. J. 2023, AJ, 165, 189

\bibitem[R. Deschamps et al. (2015)]{desetal2015}
Deschamps, R., Braun, K., Jorissen, A., et al. 2015, AA, 577, 55

\bibitem[R. Farmer et al. (2023)]{faretal2023}
Farmer, R., Laplace, E., Ma, J-z, de Mink, S. E., \& Justham, S. 2023, 
ApJ, 948, 111

\bibitem[J. D. Fernie \& N. R. Evans (1997)]{ferandeva1997}
Fernie, J. D., \& Evans, N. R. 1997, PASP, 109, 541

\bibitem[A. Fresa (1957)]{fre1957}
Fresa, A. 1957, AJ, 62, 362

\bibitem[GAIA Collaboration (2023)]{gai2023}
GAIA Collaboration, Vallenari, A., Brown, A. G. A., et al. 2023, A\&A, 674, 1

\bibitem[L. V. Glazonova et al. (2008)]{glaetal2008}
Glazunova, L. V., Yushchenko, A. V., Tsymbal, V. V., et al. 2008, AJ, 136, 1736

\bibitem[R. O. Gray \& R. F. Garrison (1989)]{graandgar1989}
Gray, R. O., \& Garrison, R. F. 1989, ApJS, 70, 623

\bibitem[P. Harmanec (1990)]{har1990}
Harmanec, P. 1990, A\&A, 237, 91

\bibitem[W. D. Heintz (1969)]{hei1969}
Heintz, W. D. 1969, JRASC, 63, 275

\bibitem[S.-S. Huang (1963)]{hua1963}
Huang, S.-S. 1963, ApJ, 138, 342

\bibitem[I. Iben Jr. \& M. Livio (1993)]{ibeandliv1993}
Iben, I. Jr., \& Livio, M. 1993, PASP, 105, 1373

\bibitem[P. Kalv et al. (2005)]{kaletal2005}
Kalv, P., Harvig, V., Aas, T., \& Pustylnik, I. 2005, OAP, 18, 61

\bibitem[P. Kalv et al. (2009)]{kaletal2009}
Kalv, P., Aas, T., \& Harvig, V. (2009), Publications of the Tallinn 
Observatory, 6, 1

\bibitem[B. Kloppenborg et al. (2010)]{kloetal2010}
Kloppenborg, B., Stencel, R., Monnier, J. D., et al. 2010, Nature, 464, 870

\bibitem[M.-Y. Li et al. (2023]{lietal2023}
Li, M.-Y., Qian, S.-B., Zhu, L.-Y. et al. 2023, ApJS, 266, 28

\bibitem[A. Mainzer et al. (2014)]{maietal2014}
Mainzer, A., Bauer, J., Cutri, R. M. et al. 2014, ApJ, 792, 30

\bibitem[E. F. Milone (1968)]{mil1968}
Milone, E. F. 1968, AJ, 73, 708 

\bibitem[D. Monin et al. (2014)]{monetal2014}
Monin, D., Saddlemyer, L., \& Bohlender, D. 2014, RevMex, 45, 69

\bibitem[D. Mourard et al. (2018)]{mouetal2018}
Mourard, D., Broz, M., Nemravova, A., et al. 2018, A\&A, 616, A112

\bibitem[C. A. Nelson \& P. P. Eggleton (2001)]{nelandegg2001}
Nelson, C. A., \& Eggleton, P. P. 2001, ApJ, 552, 664

\bibitem[G. Neugebauer et al. (1984)]{neuetal1984}
Neugebauer, G., Habing, H. J., van Duinen, R. et al. 1984, ApJ, 278, 1

\bibitem[D. J. K. O'Connell (1951)]{oco1951}
OConnell, D. J. K. 1951, Pub. Riverview College Obs., 2, 85

\bibitem[K. Pavlovski et al. (2023)]{pavetal2023}
Pavlovski, K., Southworth, J., Tkachenko, A., Van Reeth, T., \& Tamajo, E. 2023, A\&A, 671, A139

\bibitem[Z. Penoyre et al. (2022)]{penetal2022}
Penoyre, Z., Belokurov, V., \& Evans, N. W. 2022, MNRAS, 513, 5270

\bibitem[M. Plavec \& R. H. Koch (1978)]{plaandkoc1978}
Plavec, M., \& Koch, R. H. 1978, IBVS, 1482

\bibitem[D. M. Popper (1977)]{pop1977}
Popper, D. M. 1977, PASP, 89, 315

\bibitem[A. Prsa \& T. Zwitter (2005)]{prsandzwi2005}
Prsa, A., Zwitter, T. 2005, ApJ, 628, 426

\bibitem[I. Pustylnik et al. (2007)]{pusetal2007}
Pustylnik, I., Kalv, P., Harvig, V., \& Aas, T. 2007, AApT, 26, 339

\bibitem[E. H. Richardson (1968)]{ric1968}
Richardson, E. H. 1968, JRASC, 62, 313

\bibitem[J. E. Rodriguez et al. (2016)]{rodetal2016}
Rodriguez, J. E., Stassum, K. G., Lund, M. B. et al. 2016, AJ, 151, 123

\bibitem[S. Rucinski (1969)]{ruc1969}
Rucinski, S. 1969, Acta A., 19, 4

\bibitem[D. Ruzdjak et al. (2009)]{ruzetal2009}
Ruzdjak, D., Bozic, H., Harmanec, P. et al. 2009, A\&A, 506, 1319
 
\bibitem[N. N. Samus et al. (2017)]{sametal2017}
Samus, N. N., Kazarovets, E. V., Durlevich, O. V., et al. 2017, ARep, 61, 80

\bibitem[H. Sana et al. (2012)]{sanetal2012}
Sana, H., de Mink, S. E., de Koter, A., et al. 2012, Science, 337, 444

\bibitem[C.-Y. Shao \& S. Gaposchkin (1962)]{shaandgap1962}
Shao, C.-Y., \& Gaposchkin, S. 1962, AJ, 67, 283

\bibitem[M. Simon (1997)]{sim1997}
Simon, M. 1997, ApJ, 482, 81

\bibitem[M. F. Skrutskie et al. (2003)]{skretal2003}
Skrutskie, M. F., Cutri, R. M., Stiening, R. et al. 2003, ipac.data, I2S

\bibitem[A. R. Taylor et al. (2003)]{tayetal2003}
Taylor, A. R., Gibson, S. J., Peracaula, M. et al. 2003, AJ, 125, 3145

\bibitem[G. Thiessen (1956)]{thi1956}
Thiessen, G. 1956, ZfA, 39, 65

\bibitem[L. Wang et al. (2023)]{wanetal2023}
Wang, L., Gies, D. R., Peters, G. J., \& Han, Z. 2023, AJ, 165, 203

\bibitem[R. E. Wilson (1979)]{wil1979}
Wilson, R. E. 1979, ApJ, 234, 1054

\bibitem[R. E. Wilson (1993)]{wil1993}
Wilson, R. E. 1993, ASP Conference Series 38, 91

\bibitem[R. E. Wilson \& E. J. Devinney (1971)]{wilanddev1971}
Wilson, R. E., \& Devinney, E. J. 1971, ApJ, 166, 605

\bibitem[E. L. Wright et al. (2010)]{wrietal2010}
Wright, E. L., Eisenhardt, P. R. M., Mainzer, A. K., et al. 2010, AJ, 140, 1868

\bibitem[M. Xue et al. (2016)]{xueetal2016}
Xue, M., Jiang, B. W., Gao, J., liu, J., Wang, S., \& Li, A. 2016, ApJS, 224, 23

\bibitem[H. B. Yuan. et al. (2013)]{yuaetal2013}
Yuan, H. B., Liu, X. W., \& Xiang, M. S. 2013, MNRAS, 430, 2188

\bibitem[M. Zeilik et al. (1982)]{zeietal1982}
Zeilik, M., Heckert, P., Henson, G., \& Smith, P. 1982, AJ, 87, 1304

\end{thebibliography}
\end{document}